\documentclass[camera,letterpaper,nomarginnotes,nonarrowgutter]{jpaper}

\usepackage{amsmath,amssymb,amsfonts}
\usepackage{graphicx}
\usepackage{textcomp}
\usepackage{xcolor}
\usepackage{fancyhdr}
\usepackage[export]{adjustbox}
\usepackage{algorithm}
\usepackage{algorithmicx}
\usepackage{algpseudocode}

\usepackage{soul}
\usepackage{arevmath}
\usepackage{setspace}
\usepackage{xspace}
\usepackage[utf8]{inputenc}
\usepackage{blindtext, subfig}
\usepackage{fancyhdr}
\usepackage{hyperref}
\usepackage[normalem]{ulem}
\usepackage{datetime}

\usepackage{balance}

\usepackage{circledsteps}

\usepackage[compress,sort]{cite}

\newcommand{\algcomment}[1]{\textcolor{blue}{//#1}}

\newcommand{\definition}[1]{\emph{#1}}

\newcommand{\titleShort}[0]{Morpheus\xspace}
\newcommand{\cachemode}[0]{cache mode\xspace}

\newcommand{\computemode}[0]{compute mode\xspace}

\newcommand{\hkernel}[0]{extended LLC kernel\xspace}

\begin{document}
\bstctlcite{IEEEexample:BSTcontrol}
\title{\huge \titleShort: Extending the Last Level Cache Capacity \\in GPU Systems Using Idle GPU Core Resources}

\newcommand{\affilETH}[0]{\textsuperscript{\S}}
\newcommand{\affilsharif}[0]{\textsuperscript{$\dagger$}}
\newcommand{\affilipm}[0]{\textsuperscript{$\ddagger$}}
\newcommand{\affilPostec}[0]{\textsuperscript{$\nabla$}}

\author{
{*Sina Darabi\affilsharif}~~~%
{*Mohammad~Sadrosadati\affilETH}~~~%
{Joël~Lindegger\affilETH}~~~%
{Negar~Akbarzadeh\affilsharif}~~~%
{Mohammad~Hosseini\affilipm}\\%
{Jisung~Park\affilETH\affilPostec}~~~%
{Juan~Gómez-Luna\affilETH}~~~%
{Hamid~Sarbazi-Azad\affilsharif\affilipm}~~~~
{Onur Mutlu\affilETH}
\vspace{2mm}\\%
\emph{{\affilsharif Sharif University of Technology~~~~~ 
\affilETH ETH Z{\"u}rich~~~~~}}\\ \emph{{\affilipm  Institute for Research in Fundamental Sciences (IPM)~~~~~\affilPostec POSTECH}%
}%
\thanks{Sina Darabi and Mohammad Sadrosadati are co-primary authors.}
}

\maketitle
\thispagestyle{plain}

\begin{abstract}
Graphics Processing Units (GPUs) are widely-used accelerators for data-parallel applications. In many GPU applications, GPU memory bandwidth bottlenecks performance, causing underutilization of GPU cores. Hence, disabling many cores does \emph{not} affect the performance of memory-bound workloads.  %
While simply power-gating unused GPU cores would save energy, prior works  attempt to better utilize GPU cores 
for other applications (ideally compute-bound), which increases the GPU's total throughput.

In this paper, we introduce \definition{\titleShort}, a new %
hardware/software co-designed technique to boost the performance of memory-bound applications. The key idea of \titleShort~is to exploit unused core resources to extend the GPU last level cache (LLC) capacity. In \titleShort, each GPU core has two execution modes: \definition{\computemode} and \definition{\cachemode}. Cores in \definition{\computemode} operate conventionally and run application threads. However, for the cores in \definition{\cachemode}, \titleShort invokes a software helper  %
kernel that uses the cores' on-chip memories (i.e., register file,  shared memory, and L1) in a way that extends the LLC capacity for a running memory-bound workload. %
\titleShort adds a %
controller to the GPU hardware to forward LLC requests to either %
the conventional LLC (managed by hardware) or the extended LLC (managed by the helper kernel). %
Our experimental results show that \titleShort improves the performance and energy efficiency of a baseline GPU %
architecture by an average of 39\% and 58\%, 
respectively, across %
several  
memory-bound workloads. %
\titleShort' performance is within 3\% of a GPU design that has a quadruple-sized conventional LLC. %
\titleShort can thus contribute to reducing the hardware dedicated to a conventional LLC by exploiting idle cores' on-chip memory resources as additional cache capacity.%
\end{abstract}


\section{Introduction}

Graphics Processing Units (GPUs) are widely-used accelerators, especially  for data-parallel applications with high arithmetic intensity (i.e., arithmetic instructions
executed per byte accessed from memory). GPUs rely on managing execution resources for a large number of Single-Program-Multiple-Data (SPMD) threads to exploit this arithmetic intensity and
overlap the long memory access latencies with computation~\cite{chen2014adaptive,nematollahi2018neda,nematollahi2020efficient}.

Unfortunately, the maximum performance of a GPU is often 
limited by the available memory bandwidth~\cite{vijaykumar2015case}, causing considerable underutilization of GPU cores, i.e., GPU cores are frequently idle (waiting for memory accesses)  %
during application execution time. %
This is the case for many important general-purpose GPU applications, such as \texttt{kmeans}~\cite{che2009rodinia}, \texttt{mri-gri}~\cite{stratton2012parboil}, and \texttt{cfd}~\cite{che2009rodinia}%
, which are memory-bound in nature due to their low arithmetic intensity~\cite{stratton2012parboil,che2009rodinia }. 
As a result, we do \emph{not} need to use all available GPU cores to saturate the performance of these applications.
To demonstrate this, we experimentally study %
(using an NVIDIA RTX~3080 GPU~\cite{3080-white} and Accel-Sim~\cite{khairy2020accel}) the performance of 14 memory-bound applications as we scale the number of GPU cores.
Our experiments reveal that, on average across the 14 applications, %
only 56\% of the GPU cores %
are enough to saturate performance (see \S\ref{sec:motiv} for more details). 
Hence, the remaining 44\% of the GPU cores, on average, can remain \emph{unused} (i.e., no threads %
scheduled onto them) without hurting performance of memory-bound applications. %

Several prior works~(e.g., \cite{zhu2016onac,tan2012rise,wadden2014real,themis,6742976,6168946,8327010,8409306,jog2016exploiting,narasiman2011improving}) make similar observations and propose to have \emph{only} a \emph{subset} of the available GPU cores execute memory-bound application threads, and leverage the remaining GPU cores %
in one of three ways: (1)  \emph{power-gating} these cores to save energy~\cite{zhu2016onac}, (2) using them for \emph{redundant execution} of the \emph{already running} memory-bound application for better reliability~\cite{tan2012rise,wadden2014real}, and (3) \emph{co-scheduling} additional \emph{compute-bound} applications to these cores to increase GPU's total throughput~\cite{themis,6742976,6168946,8327010,8409306,jog2016exploiting,narasiman2011improving}.
In contrast, \textbf{our goal} is to boost the performance of a %
memory-bound application using a number of GPU cores that are not useful for executing application threads. 

To this end, we propose \emph{\textbf{\titleShort}}.\footnote{\textit{``\titleShort''} because we dynamically \emph{morph}, parts of GPU hardware (in a logical sense) to meet memory-bound applications' needs.} %
\textbf{The key idea of \titleShort{}} is to reserve a number of GPU cores to
use their on-chip memory (i.e., register files, shared memory, and L1 cache) as an extension of the shared last level cache (LLC, e.g., the shared L2 cache in NVIDIA GPUs). Doing so can improve the performance of memory-bound applications %
due to two main reasons. First, a larger LLC reduces the number of off-chip memory accesses, thereby enabling more threads to run effectively since the system is not memory bandwidth bottlenecked any more. Second, a larger LLC reduces memory access latency, %
 which can improve performance.

\titleShort introduces two  \definition{execution modes} (compute and cache) for every GPU core. A core in \definition{\computemode} behaves like a regular core in conventional GPUs, i.e., it executes application threads. %
A core in \definition{\cachemode} %
lends its on-chip memory %
space to extend the effective shared LLC size via a hardware/software co-designed technique.  %
A GPU core in \cachemode runs a software helper kernel, called the \emph{\hkernel}, that  stores the extended LLC tag/data arrays inside the GPU core's local register file, shared memory, and L1 cache. We add a hardware controller, called the \emph{\titleShort{} controller}, to support access to the extended LLC. The \titleShort{} controller performs three main tasks: it %
(1)~forwards each LLC request to either the conventional LLC or the extended LLC, depending on the requested memory address (i.e., LLC set number), (2)~tracks outstanding extended LLC requests, and (3)~predicts the outcome of an extended LLC lookup (hit/miss), so that it forwards \emph{only} the requests that are predicted to be hit in the extended LLC to GPU cores in \cachemode, which mitigates the overhead of extended LLC misses. The \titleShort{} controller uses Bloom filters~\cite{bloom1970space} for hit/miss prediction, providing \emph{zero} false-negative and negligible false-positive rates.

To improve the effectiveness of \titleShort{}, we employ two optimization techniques on top of our basic design. First, we increase the effective capacity of the extended LLC by employing a cache  compression technique in the \hkernel.
Second, we accelerate the data array access in the extended LLC by adding a new specialized instruction to the GPU instruction set architecture. This new instruction enables \emph{indirectly} addressing a register, i.e., reading from a register whose index is determined by accessing the value in another register. %

To evaluate the effectiveness of \titleShort{}, we first measure 
the bandwidth, access latency, and energy consumption of the extended LLC %
via real-system experiments using an NVIDIA RTX~3080 GPU (\S\ref{sec:SC-study}).  %
We then use the measured performance and energy numbers for the extended LLC in the AccelSim simulator~\cite{khairy2020accel} to estimate the effect of \titleShort{} on overall GPU performance and energy consumption. Our experimental results show that %
\titleShort{} improves overall GPU performance and energy
efficiency by an average of 39\% and 58\%, respectively, compared to the baseline NVIDIA RTX~3080 GPU architecture, across 14 memory-bound applications. \titleShort{} performs within 3\% of a conventional GPU design that has a %
quadruple-sized conventional LLC (assuming \emph{no} latency overhead for the larger conventional LLC).

\vspace{-0.5mm}
We make the following contributions:
\vspace{-0.5mm}
\begin{itemize}
    \item We propose \titleShort{}, the first technique %
    that leverages some %
    GPU cores' on-chip memories to extend the total GPU last-level cache capacity.
    \item We introduce two new mechanisms that \titleShort{} employs: (1) a software helper kernel to extend the LLC using the on-chip memory resources of GPU cores that are in \cachemode, %
    and (2) a hardware controller that enables accesses to \emph{both} the conventional LLC and the extended LLC. 
    \item We evaluate \titleShort and show that it significantly improves performance and energy. It also enables a 4$\times$ larger cache without requiring new hardware resources for it. %
\end{itemize}

\section{Background}
\label{sec:back}

A GPU program is composed of a number of kernels that are executed using Single-Program-Multiple-Data (SPMD) threads~\cite{PROGRAMGUIDE}. These threads are partitioned into multiple blocks,  or Cooperative Thread Arrays (CTAs)~\cite{PROGRAMGUIDE}. CTAs are then assigned to Single-Instruction-Multiple-Data (SIMD) cores for execution, called \definition{Streaming Multiprocessors (SMs)} on NVIDIA GPUs. Each SM contains multiple CUDA cores, special-function units (e.g., cos, sin and tan), shared memory, L1 cache and thousands of registers. %
SMs are connected to several memory partitions using an on-chip interconnection network. Each memory partition includes one or multiple LLC banks, a memory controller, and a main memory (GDDRx/HBMx).
Threads inside each CTA are grouped into \definition{warps}. %
Threads within a warp execute the same instruction on different data items in a lock-step manner.
The \definition{warp scheduler} time-multiplexes warps, and assigns warps to the execution units. %

\section{Motivation}
\label{sec:motiv}

\begin{figure*}[h]
\begin{center}
  \includegraphics[trim= 0mm 0mm 0mm 0mm,width=1\linewidth]{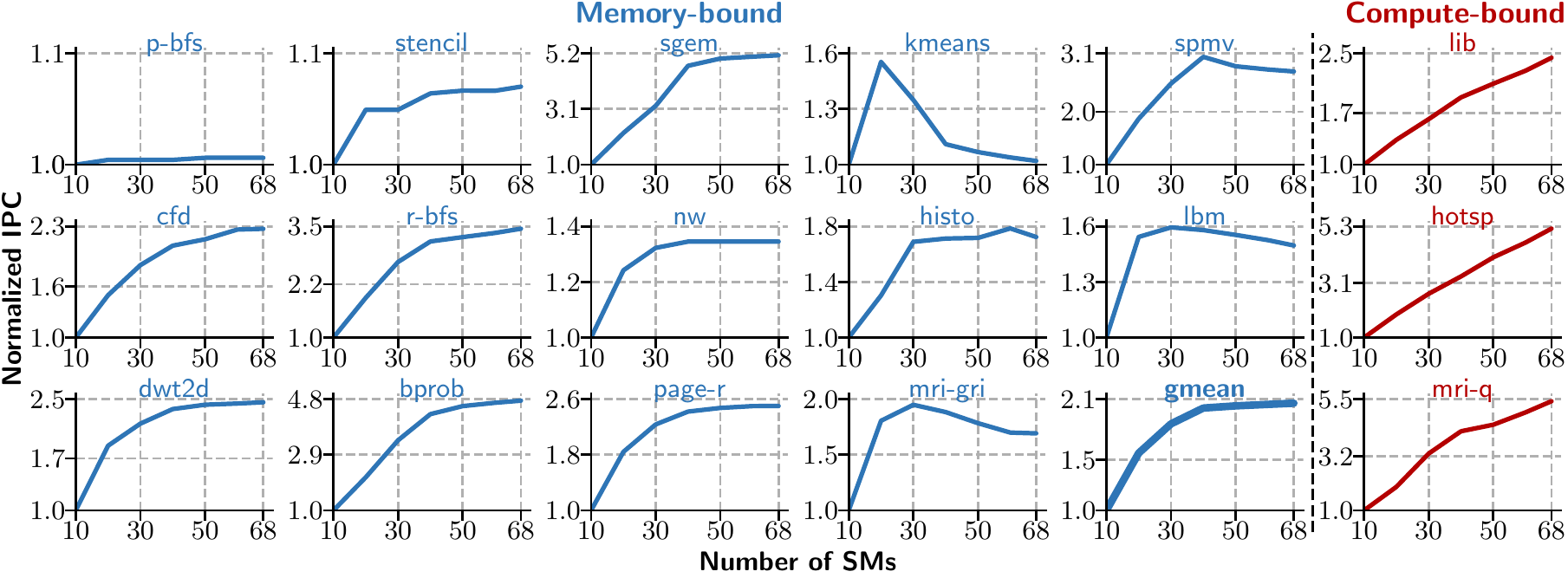}
  \caption{Normalized GPU performance (IPC) of 14 memory-bound and 3 compute-bound applications %
 }
  \label{fig:perf-sat}
\end{center}
\end{figure*}

Memory-bound applications cannot fully utilize the compute throughput of GPUs as they are bottlenecked by the limited memory bandwidth. This causes long memory access latencies, which cannot be hidden via thread-level parallelism, causing core idleness %
and performance saturation~\cite{sadrosadati2018ltrf,vijaykumar2015case,oh2019linebacker,khorasani2018register,xie2015enabling,vijaykumar2016zorua,chatterjee2017architecting,zhu2018performance,agarwal2015unlocking,zhao2012optimizing}. %
To illustrate this observation, we experimentally study the performance of 17 applications (14 memory-bound and 3 compute-bound) as we scale the number of GPU cores using AccelSim~\cite{khairy2020accel} (see \S\ref{sec:Method} for our methodology).  %
Figure~\ref{fig:perf-sat} illustrates the results for a baseline GPU architecture that resembles an NVIDIA RTX~3080~\cite{3080-white}. %
The x-axes correspond to the number of GPU cores~(SMs), ranging from 10 (chosen empirically) to 68 (the total SM count in an RTX~3080). %
The y-axes correspond to the normalized performance for each application. We normalize the performance of each application to its performance when using 10 SMs for readability.  %

We make two key observations. First, performance \emph{gradually saturates} (i.e., stops increasing) as the number of SMs increases for 9 of the memory-bound applications (\texttt{p-bfs}, \texttt{cfd}, \texttt{dwt2d}, \texttt{stencil}, \texttt{r-bfs}, \texttt{bprob}, \texttt{sgem}, \texttt{nw}, \texttt{page-r}). In contrast, the performance of the compute-bound applications continues to increase with more SMs.
Second, performance \emph{decreases} sharply after a certain number of SMs for 5 of the memory-bound applications (\texttt{kmeans}, \texttt{histo}, \texttt{mri-gri}, \texttt{spmv}, \texttt{lbm}). 
For example, after more than 20 SMs are used for the kmeans application, performance drops greatly and the performance of the application with 68 SMs is almost the same as it is with 10 SMs, which is 50\% lower than with 20 SMs.
We conclude that limiting the number of SMs running a memory-bound application (after some SM count) not only does \emph{not} significantly hurt performance but can even \emph{improve} performance in some cases. %

\begin{figure}[h]
\begin{center}
  \includegraphics[trim= 0mm 0mm 0mm 0mm,width=1\linewidth]{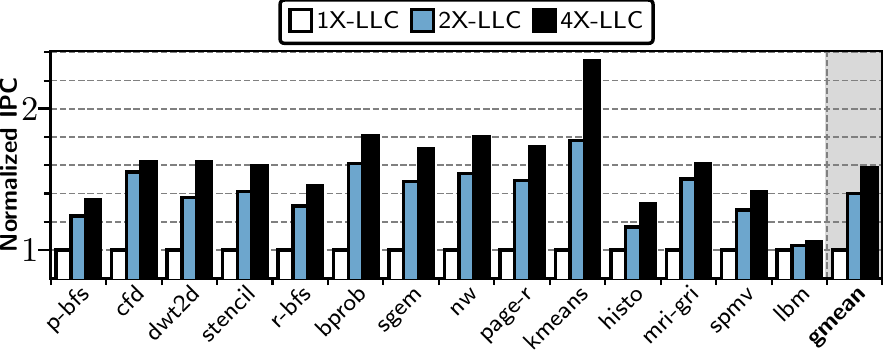}
  \caption{Effect of larger LLC sizes (2$\times$ and 4$\times$ the size of the baseline LLC) on performance (normalized IPC relative to the baseline LLC) of 14 memory-bound applications}
  \label{fig:perf-sat-2x}
\end{center}
\end{figure}

Leveraging our key observations, we aim to reserve a number of SMs and use  %
their on-chip memory (i.e., register file, shared memory, and L1 cache) to extend the overall LLC capacity and, thus, improve GPU performance for memory-bound applications. 
To quantify the benefits of potentially having a larger LLC for memory-bound applications, we repeat our previous experiment with 2$\times$ and 4$\times$ LLC sizes, compared to the baseline 5-MiB LLC of an NVIDIA RTX~3080. We vary the number of GPU cores in these two evaluated GPU designs (2$\times$ and 4$\times$ larger LLCs) %
and measure overall performance for each case. Figure~\ref{fig:perf-sat-2x} shows the maximum performance that we observe while varying the number of GPU cores for %
14 representative memory-bound applications. We normalize the performance results of each application to the case of a baseline 5-MiB LLC. We observe that both 2$\times$ and 4$\times$  LLC sizes improve the performance of \emph{all} evaluated memory-bound applications. In particular, a 4$\times$ larger LLC improves performance by as much as 2.34$\times$ (kmeans), and by 1.57$\times$ on average (geometric mean). We conclude that a larger LLC effectively and consistently improves the performance of memory-bound applications.

\textbf{Our goal} in this work is to design a mechanism that can leverage the on-chip memory units of a number of GPU cores (that are otherwise not beneficial), to effectively extend the overall LLC capacity for memory-bound GPU applications, so as to increase performance.

\section{\titleShort}
 \label{sec:Mech}
 
We introduce \titleShort, a hardware/software co-designed technique %
to alleviate the memory bottleneck on GPUs. The key idea of \titleShort is to reserve a number of GPU cores and use their on-chip memory units (i.e., register file, shared memory, and L1
cache) as an extension of the GPU's LLC. 
In \titleShort, each GPU core has two working modes, \definition{\computemode}~and \definition{\cachemode}. Cores in \computemode behave exactly like the cores in existing GPUs and execute application threads. In contrast, cores in \cachemode~execute a software helper kernel (called the \emph{\hkernel}) to extend the LLC capacity by exploiting the storage capacity of their local on-chip memory units. This additional LLC capacity provided by the cores in \cachemode is called the \emph{extended LLC}. 

\pgfkeys{/csteps/inner color=white}
\pgfkeys{/csteps/fill color=black}

\begin{figure}[h]
\begin{center}
  \includegraphics[width=1\columnwidth]{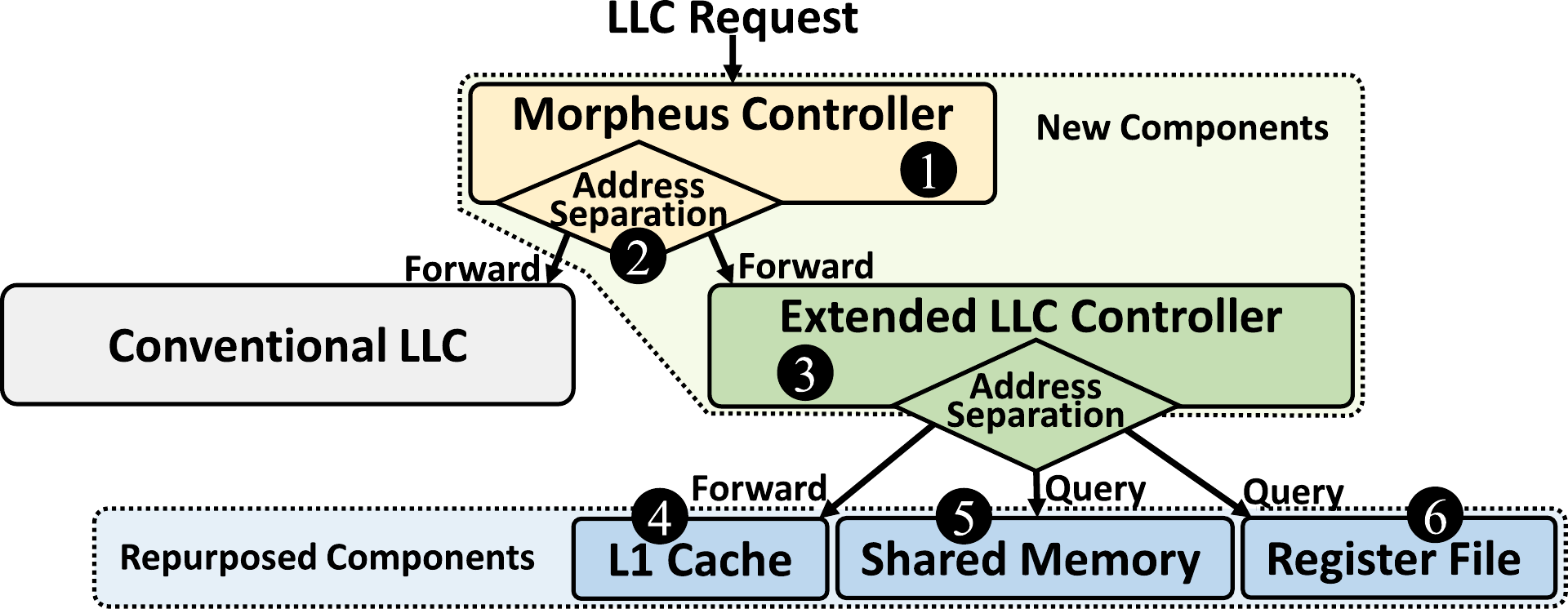}
  \caption{Conceptual overview of \titleShort} %
  \label{fig:morpheus-overview}
\end{center}
\end{figure}

Figure~\ref{fig:morpheus-overview} gives a conceptual overview of the  LLC lookup procedure in \titleShort{}. An LLC request in a \titleShort-enabled GPU first arrives at the \emph{\titleShort{} controller}~(\Circled{1} in Figure~\ref{fig:morpheus-overview}). The \titleShort controller forwards the LLC request to either the conventional LLC (which works exactly as in existing GPUs) or the \emph{extended LLC} (which we propose). The forwarding decision is based on a static address mapping scheme, called \emph{address separation}~\Circled{2}. An access to the extended LLC is served by the \emph{extended LLC controller}~\Circled{3} using the on-chip memory units of cores in \cachemode. The extended LLC controller either \emph{forwards} a given request to the core's L1 cache~\Circled{4}, or \emph{directly queries} the register file~\Circled{5} or shared memory~\Circled{6}. The forwarding or querying decision is based on the same static address separation principle as in the \titleShort controller. %

\pgfkeys{/csteps/inner color=black}
\pgfkeys{/csteps/outer color=black}
\pgfkeys{/csteps/fill color=white}

\begin{figure}[h]
\begin{center}
  \includegraphics[trim= 0mm 0mm 0mm 0mm,width=1\columnwidth]{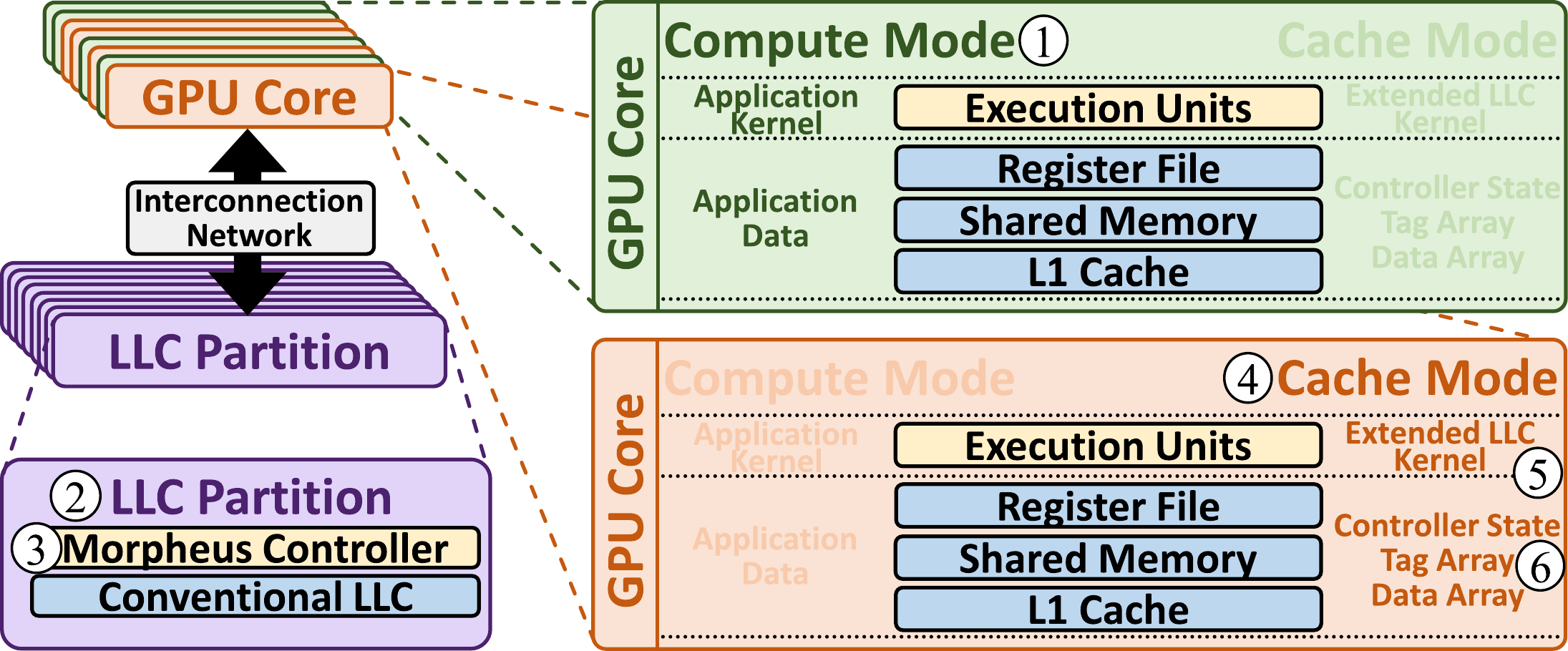}
  \caption{GPU structure with \titleShort} %
  \label{fig:proposed-GPU}
\end{center}
\end{figure}

\titleShort implements the \titleShort controller as a new hardware unit per LLC partition and the extended LLC controller as software (i.e., the \hkernel) running on cores in \cachemode. Figure~\ref{fig:proposed-GPU} shows a pictorial example of a \titleShort-enabled GPU with these new components. An LLC request originates from a core in \computemode (\Circled{1} in Figure~\ref{fig:proposed-GPU}) and moves through the interconnection network to an LLC partition~\Circled{2} based on a static address mapping scheme, similar to the one used in a conventional GPU. In a \titleShort-enabled GPU, a hardware implementation of the \titleShort controller local to the LLC partition~\Circled{3} then either forwards the request to the local conventional LLC, or through the interconnection network to the responsible GPU core running in \cachemode~\Circled{4} based on a static address separation scheme~(\S\ref{sec:add-sep}). The responsible \cachemode GPU core is determined by the address separation scheme. The extended LLC controller is implemented as multiple instances of a software helper kernel (the \hkernel) running on cores in \cachemode~\Circled{5}. This kernel queries one of the local memory units~\Circled{6} for each incoming request, based on a static address separation scheme~(\S\ref{sec:add-sep}), and sends the response over the interconnection network back to the \titleShort controller~\Circled{3}. If the request is a hit in the extended LLC, the cache block is sent over the interconnection network to the GPU core that initially issued the LLC request~\Circled{1}. If the request is a miss in the extended LLC, it is treated exactly like an LLC miss in a conventional GPU by the LLC partition.

The rest of this section explains \titleShort{}' key mechanisms in detail. 
\S\ref{sec:morpheus} describes our hardware implementation of the \titleShort{} controller. \S\ref{sec:soft-cache} describes our software implementation of the extended LLC controller using the \hkernel. \S\ref{sec:compressor} introduces two optimization techniques on top of our basic design to improve the effectiveness of \titleShort{}.

\subsection{\titleShort{} Controller}
\label{sec:morpheus}

This unit has three main tasks: (1) separating LLC requests between the conventional LLC and the extended LLC, (2) handling communication between the extended LLC and the LLC partition, and (3) predicting the outcome of the extended LLC lookup (hit/miss), so that the \titleShort controller forwards only the requests that are predicted to be hits in the
extended LLC to GPU cores in \cachemode, which mitigates the
overhead of extended LLC misses. 
Figure~\ref{fig:Logical-LLC-Controller} shows the main components in the \titleShort{} controller. In this section, we describe each component in detail.  %

\subsubsection{Address Separation}
\label{sec:add-sep}
Since a Morpheus-enabled GPU  employs two distinct LLCs (i.e., the conventional LLC and the extended LLC), there should be a mechanism to map each cache block to one of these LLCs. To this end, \titleShort{} 
divides the memory address space statically into two partitions proportional in size to the conventional and extended LLC capacity. The conventional LLC is responsible for caching the first partition, while the extended LLC is responsible for caching the second. %
When the \titleShort{} controller receives an LLC request, a unit called \emph{address separator}  checks whether or not the set number is in the range of memory addresses served by the extended LLC. If so, the unit forwards the LLC request to the next unit in the \titleShort{} controller, the \emph{hit/miss predictor.} %
Otherwise, the conventional LLC handles the request exactly the same way as done in the conventional GPU architectures.

\subsubsection{Hit/Miss Prediction}
\label{sec:misspenalty}
\begin{figure}[h]
\begin{center}
  \includegraphics[width=\columnwidth]{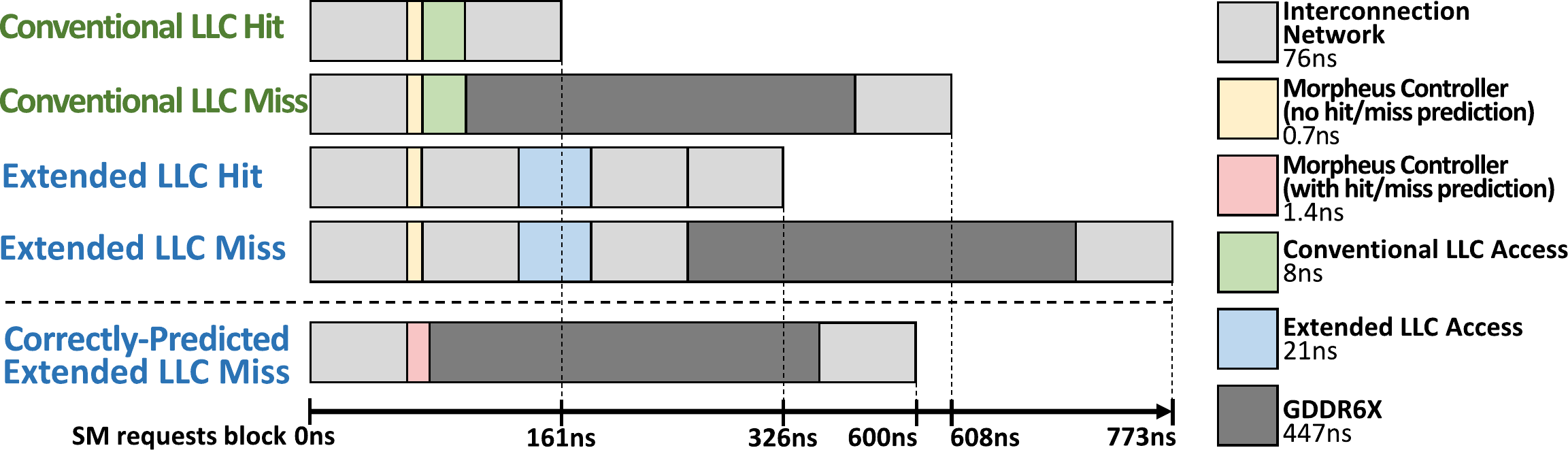}
  \caption{Timelines for LLC hits, misses, and predicted misses on a \titleShort-enabled GPU} %
  \label{fig:SW-LLC-Overhead}
\end{center}
\end{figure}

Misses in the extended LLC are more expensive in terms of latency compared to misses in the conventional LLC. Figure~\ref{fig:SW-LLC-Overhead} breaks down the latencies of hits and misses in both the conventional and extended LLC (see \S\ref{sec:Method} for our methodology). We observe that misses in the conventional LLC take 608ns to be served, while misses in the extended LLC take 773ns (i.e., 27\% longer). This is due to two  main reasons. First, an extended LLC miss adds a round trip interconnection network latency (two grey boxes in Figure~\ref{fig:SW-LLC-Overhead}) compared to a conventional LLC miss, to move the request and response between the \titleShort{} controller and a GPU core in \cachemode.
Second, the latency for accessing the extended LLC is longer than for the conventional LLC, because in the extended LLC, tag lookups and data array accesses have to be managed by the \hkernel{} (i.e., in software).

We mitigate the overhead of extended LLC misses using a \definition{hit/miss predictor}. After the address separation mechanism in the \titleShort{} controller determines that a given request falls into the extended LLC's address space, the hit/miss predictor decides if the request is likely to be a hit in the extended LLC. If the request is predicted to be a hit, the \titleShort{} controller forwards the request to the extended LLC. Otherwise, the \titleShort{} controller directly accesses the off-chip DRAM to serve the memory request. 

Figure~\ref{fig:SW-LLC-Overhead} shows the timeline for a correctly-predicted extended LLC miss. If a miss is predicted correctly, it
avoids the unnecessary latency of (1)~an interconnection network roundtrip, and (2)~querying the tag array with the \hkernel. %
Thus, correctly-predicted extended LLC misses can be serviced as fast as conventional LLC misses.

To maintain correctness in the presence of the hit/miss predictor, we must understand if and which types of mispredictions are acceptable.
We observe that it is acceptable to falsely predict a request as an LLC hit when in reality the request is a miss. This is because the predicted LLC hit causes a lookup in the extended LLC, at which point the misprediction will be discovered. Nevertheless, such \emph{false positives} are undesirable, because they increase the latency of extended LLC misses to the same latency as if there were no hit/miss prediction. %
In contrast, falsely predicting a request as an LLC miss even though the requested address is in the LLC can violate both coherence and consistency guarantees. For example, if a request to a dirty cache block is falsely predicted as an LLC miss, the requesting core will receive an out-of-date value from main memory, which violates basic cache correctness. Thus, for correctness, the hit/miss predictor must not produce such \emph{false negatives}, or otherwise it should recover from its mispredictions such that no correctness problems appear.

We design our hit/miss predictor using Bloom filters~\cite{bloom1970space}. Bloom filters are a good fit for our requirements because they provide fast and low-cost set membership queries without false negatives. In our hit/miss predictor, a Bloom filter represents the LLC blocks which are currently in a given extended LLC set.
To avoid increasingly frequent false positives, the Bloom filters must be cleared regularly,\footnote{Counting Bloom filters~\cite{fan2000summary} would support individual element removal instead, but require more bits compared to standard Bloom filters.} after which there is a risk for \emph{false negatives}. We next describe an algorithm using two Bloom filters, which are cleared alternately, thus avoiding false negatives.

\pgfkeys{/csteps/inner color=white}
\pgfkeys{/csteps/outer color=black}
\pgfkeys{/csteps/fill color=black}

The key idea of our extended LLC hit/miss prediction algorithm is to track the blocks currently in the LLC set using two Bloom filters per extended LLC set, BF1 and BF2. BF1 and BF2 are updated upon every access to the extended LLC set, such that at any given time, (1)~BF1 contains at least all the cache blocks currently in the extended LLC set, and (2)~BF2 contains the \emph{n} most recently used cache blocks in the extended LLC set. Invariant~(1) guarantees the absence of false negatives when querying BF1 with requested extended LLC addresses, i.e., BF1 can be used to safely predict if a request will hit in the extended LLC set. Invariant~(2) enables eventually replacing BF1 with BF2, when BF2 contains all cache blocks in the extended LLC set (\emph{n}$\geq$associativity). The benefit of eventually replacing BF1 with BF2 is that BF2 does not (yet) contain any evicted cache blocks, thus producing fewer false positives than the old BF1. In contrast, BF1 may contain multiple evicted cache blocks. We explain in detail how the hit/miss predictor queries BF1, and how BF1 and BF2 are updated upon every access to the extended LLC set to maintain invariants (1) and (2).

\begin{figure}[]
\begin{center}
  \includegraphics[width=1\columnwidth]{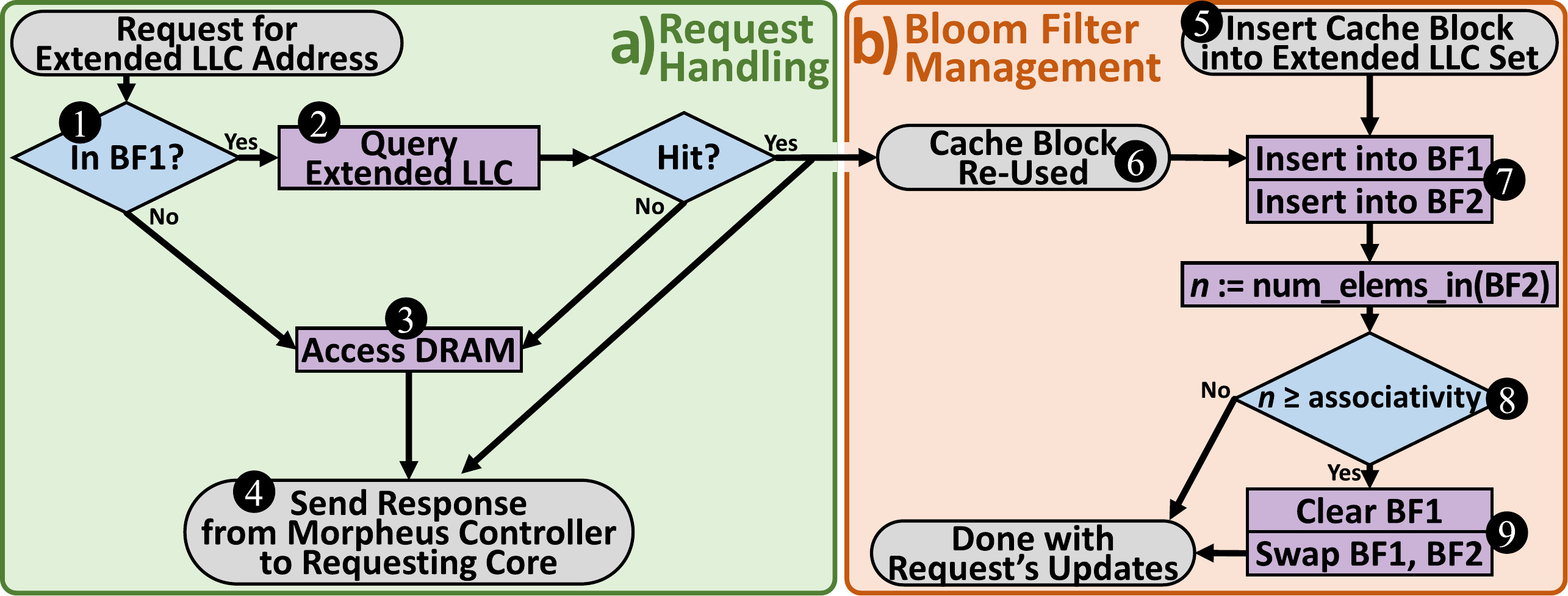}
 \caption{Flowchart of the extended LLC hit/miss predictor} %
  \label{fig:BloomFilter}
\end{center}
\end{figure}

Figure~\ref{fig:BloomFilter}(a) shows a flow diagram of how the \titleShort controller with the hit/miss predictor serves an extended LLC request. To make a hit/miss prediction for some LLC request, the hit/miss predictor queries BF1 with the LLC request's address (\Circled{1} in Figure~\ref{fig:BloomFilter}). It predicts a hit when the address is found in BF1, and a miss otherwise. This cannot produce false negatives, since invariant~(1) guarantees that BF1 contains all blocks currently in the extended LLC set, and Bloom filters do not produce false negatives. Upon a predicted extended LLC hit, the algorithm queries the address in the extended LLC~\Circled{2}. Upon a (predicted or actual) extended LLC miss, the \titleShort controller accesses the requested block in DRAM instead~\Circled{3}. Finally, the response is sent from the \titleShort controller to the core that issued the request~\Circled{4}.

Figure~\ref{fig:BloomFilter}(b) shows a flow diagram of how the hit/miss predictor updates BF1 and BF2 for a given extended LLC access to maintain invariants~(1) and~(2). Upon an access (i.e., when a cache block is inserted into the set~(\Circled{5} in Figure~\ref{fig:BloomFilter}), or a cache block in the set is re-used~\Circled{6}), the accessed cache block is inserted into both Bloom filters~\Circled{7}. This trivially maintains invariant~(1), because any inserted extended LLC block will also be in BF1. Invariant~(2) is maintained because before the insertion, BF2 contained the \emph{n}$_{old}$ most recently used blocks, and after the insertion BF2 contains either the \emph{n}=\emph{n}$_{old}$+1 most recently used blocks (if the used block was not in BF2 before) or the \emph{n}=\emph{n}$_{old}$ most recently used blocks (if the used block was already in BF2). After \emph{n} becomes as large as the set's associativity~\Circled{8}, all future predictions can be made by querying BF2 instead of BF1, without risking false negatives. This is because the extended LLC uses the LRU (least recently used) replacement policy and BF2 contains the \emph{n} most recently used blocks, thus if \emph{n} is as large as the set's associativity, BF2 is guaranteed to contain \emph{all} blocks that are in the LLC set. Thus, the contents of BF1 are cleared, BF1 and BF2 are swapped, and the scheme repeats~\Circled{9}.

\noindent\textbf{Cost.} Assuming each Bloom filter is 32 bytes in size, and up to 256 extended LLC sets per LLC partition, the hit/miss predictor requires 32B$\times$2$\times$256=16KiB of Bloom filter storage in each LLC partition.

\subsubsection{Extended LLC Query Logic Unit}
\label{sec:warp-status}
\begin{figure}[]
\begin{center}
  \includegraphics[trim = 20mm 0 20mm 0,width=0.8\linewidth]{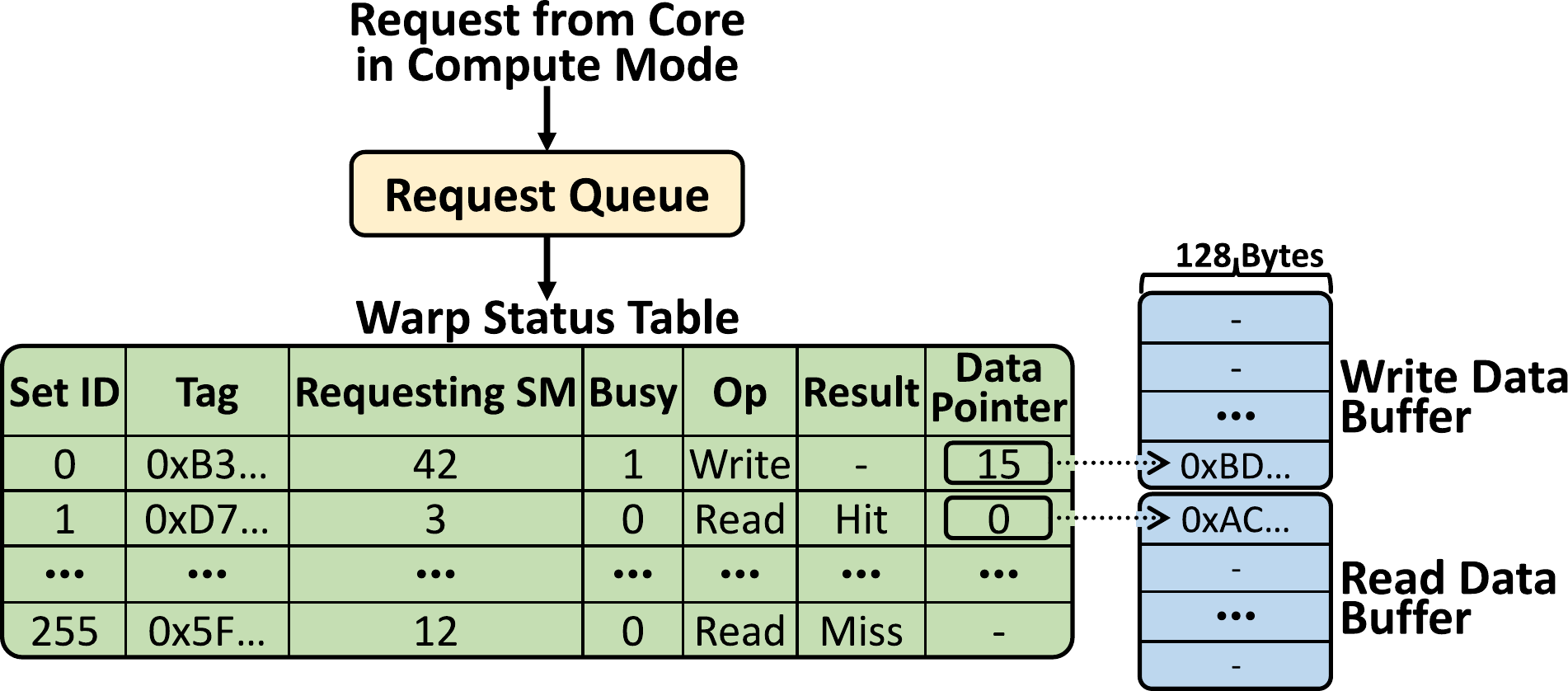}
  \caption{Extended LLC Query Logic Unit}
  \label{fig:Logical-LLC-Controller}
\end{center}
\end{figure}

The \titleShort{} controller includes a hardware unit called the \emph{extended LLC query logic unit} to track and manage outstanding extended LLC requests. Figure~\ref{fig:Logical-LLC-Controller} shows the four main components in the extended LLC query logic unit, namely the \emph{request queue}, the \emph{warp status table}, the \emph{read data buffer}, and the \emph{write data buffer}. We explain each component in this section.\\
\textbf{Request Queue.}
For simplicity, each \hkernel warp serves only a single extended LLC request at a time in \titleShort. To avoid clogging the interconnection network with backlogged request bursts, we introduce a request queue, which buffers any requests to the extended LLC. A given request is de-queued as soon as the warp assigned to the request's extended LLC set is ready to serve a new request. When de-queued, the request's metadata is written to the corresponding row in the warp status table. If the request is a write, the payload data is written to the write data buffer.\\
\textbf{Warp Status Table.}
The warp status table has one row per set in the current LLC partition, tracking the status of the \hkernel warp that is assigned to each set. Each row has fields for the current request's tag, the origin of the request, a \emph{busy} bit (indicating if the warp is currently serving a request), an \emph{op} field (indicating if the request is a read or write), a \emph{result} field (indicating if the request is a hit or a miss in the extended LLC), and a pointer to either the read or write data buffer entry for the payload data. The warp status table is memory-mapped, thus the \hkernel warps can read from and write to it with regular load/store instructions.

We size the warp status table based on the maximum number of sets in the extended LLC. We assume an NVIDIA RTX~3080 GPU as the baseline, which has 10 LLC partitions, 68 SMs, and 48 warps per SM (and thus, up to 48 extended LLC sets per SM in \cachemode). We assume that up to 75\% of all SMs can be in \cachemode, based on Figure~\ref{fig:perf-sat}, where 
GPU performance starts to saturate 
after using at least 25\% of the SMs for computation. Under these assumptions, \titleShort can provide up to 2448 extended LLC sets, i.e., about 256 sets per LLC partition. As a result, the warp status table has 256 rows in each \titleShort controller of a \titleShort-enabled NVIDIA RTX~3080 GPU.\\
\textbf{Read and Write Data Buffers.}
The read and write data buffers hold payload data from requests to the extended LLC. For example, when a write request arrives at the extended LLC query logic unit, it includes a dirty cache block (e.g., 128 bytes) to write to the extended LLC. These 128 bytes are written to an entry in the write data buffer, and the data pointer in the warp status table is updated to point to that entry. Read requests work analogously, with the difference that the \hkernel writes the requested cache block to the read data buffer entry instead. Like the warp status table, the read and write data buffers are memory-mapped, and thus \hkernel warps can read from and write to them with simple load/store instructions.

\subsection{Extended LLC Controller}
\label{sec:soft-cache}
\titleShort{} implements the extended LLC controller as a software helper kernel, called the \emph{\hkernel}, that performs three main tasks: (1)~storing and updating the extended LLC tags and data in the local memory units of a GPU core operating in \cachemode~(\S\ref{sec:SC-RF}-\ref{sec:SC-L1}), (2)~executing simple operations (e.g., increment) on the extended LLC blocks needed for \emph{atomic} instructions~(\S\ref{sec:supporting_atomics}), and (3)~identifying the correct memory unit to perform these operations on, based on a static address separation mechanism.

\titleShort{} schedules one copy of the \hkernel on each GPU core that is operating in \cachemode. The \hkernel uses the maximum number of warps in each GPU core, and each warp handles \emph{exactly} one extended LLC set. The \hkernel~effectively uses all of the local memory units of GPU cores operating in \cachemode to store the extended LLC data and metadata (e.g., tags, valid bits, dirty bits), except for a number of registers that the \hkernel reserves as \emph{auxiliary registers} for its own operation. In this section, we explain only the first and second tasks of the \hkernel~since the third task, i.e., address separation, is analogous to the address separation in the \titleShort{} controller~(\S\ref{sec:add-sep}), with the only difference being that the address space is divided proportionally to the respective \emph{memory units'} capacities instead.

\pgfkeys{/csteps/inner color=black}
\pgfkeys{/csteps/outer color=black}
\pgfkeys{/csteps/fill color=white}

\subsubsection{Extended LLC via Register File}
\label{sec:SC-RF}
In this section, we describe how  the \hkernel (1)~lays out the extended LLC blocks' tags and data in the register file, and (2)~accesses and updates them.

\begin{figure}[h]
\begin{center}
  \includegraphics[trim={0mm 0mm 8mm 0mm},width=1\linewidth]{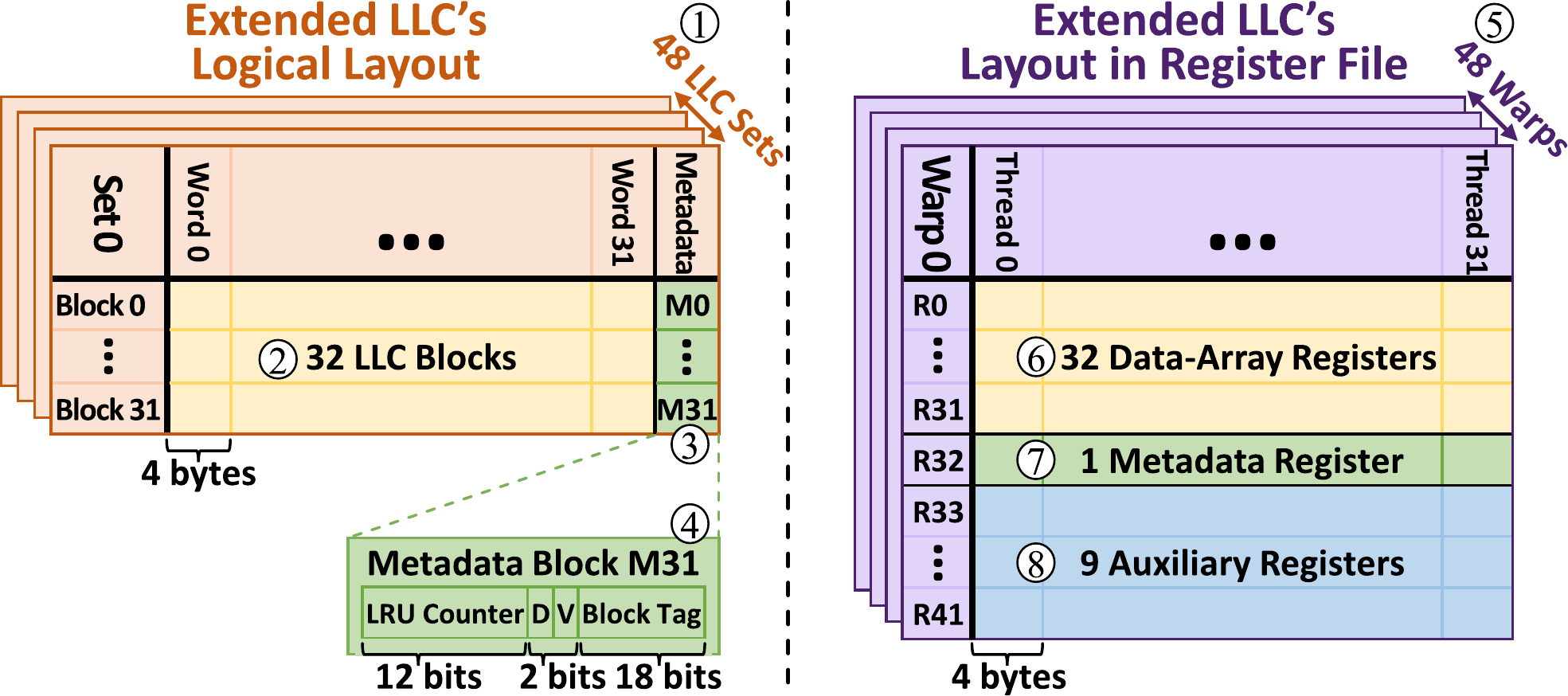}
  \caption{Each out of the 48 warps in an SM in \cachemode implements one fully associate cache set (e.g., set 0) with several cache blocks (e.g., 32).} %
  \label{fig:set-mapping}
\end{center}
\end{figure}

\noindent\textbf{Extended LLC Layout in the Register File.} Figure~\ref{fig:set-mapping} shows how the \hkernel lays out 48 sets of a 32-way set-associative extended LLC in the register file of a GPU core operating in \cachemode. This example assumes a baseline NVIDIA RTX~3080~\cite{3080-white} GPU, with up to 48 active warps per SM, and 42 registers per warp.

The \hkernel uses the register file to store a number of extended LLC sets (e.g., 48)~\Circled{1}, for each set a number of cache blocks (e.g., 32 blocks of 128 bytes each)~\Circled{2}, and for each block a metadata block~\Circled{3} containing the block's LRU counter, dirty bit, valid bit, and tag~\Circled{4}. Each extended LLC set~\Circled{1} is stored in the registers of exactly one warp~\Circled{5}. Each data block of a set~\Circled{2} is stored in exactly one warp register, called a \emph{data-array register}~\Circled{6}. The metadata blocks~\Circled{3} are coalesced into a single warp register, called the \emph{metadata register}~\Circled{7}, such that thread \textit{i} holds block \textit{i}'s metadata block. The remaining \emph{auxiliary registers} are used for \hkernel execution~\Circled{8}.

In this configuration of 48 extended LLC sets per GPU core operating in \cachemode, 32 blocks per set, and 128 bytes per block, each GPU core operating in \cachemode adds 48$\times$32$\times$128B$=$192KiB capacity to the extended LLC by using the space in the GPU core's register file.

\noindent\textbf{Extended LLC Tag Lookup.} 
When a warp executing the \hkernel receives an extended LLC request for its set, the warp executes a tag lookup procedure to determine if the request is a hit in the set,
and if so, which cache block in the set was hit.
To this end, the warp compares the tag of the request address to the tags of extended LLC blocks stored in its metadata register $R_M$ (e.g., $R_{32}$ in Figure~\ref{fig:set-mapping}).%

Algorithm~\ref{algo:hk1} shows the pseudo code for the tag lookup procedure in the \hkernel. The procedure receives the tag of the extended LLC request as the input and returns the outcome of the lookup (``HIT=True'' and the ``BLOCK\_INDEX'' of the corresponding cache block, or ``HIT=False'') as the output. Algorithm~\ref{algo:hk1} assumes that the request's tag is in the auxiliary register $R_{aux_0}$ (e.g., $R_{33}$ in Figure~\ref{fig:set-mapping}) for all threads, i.e., there are 32 copies of the tag.

Each thread of the warp is assigned to a cache block (Block 0-31 in Figure~\ref{fig:set-mapping}) and corresponding metadata block (M0-M31 in Figure~\ref{fig:set-mapping}), such that all following operations run in parallel for all cache blocks. First, each thread ensures that its assigned metadata block in $R_{M}$ (e.g., $R_{32}$ in Figure~\ref{fig:set-mapping}) is valid by checking the valid bit (line 2). Second, each thread compares the tag in its assigned metadata block to the request's tag in $R_{aux_0}$, and stores the result in an auxiliary register (e.g., $R_{aux_1}$) (line 3). Third, the comparison result of each thread is shared among all threads as a 32-bit bitvector and written into $R_{aux_2}$ using the \emph{ballot\_sync} instruction~\cite{PROGRAMGUIDE} (line 4). If $R_{aux_2}$ is non-zero, one of the tags in the metadata blocks must have matched the request's tag, i.e., the request is a hit (lines 5-6). In this case, the 0-based index of the first 1-bit in $R_{aux_2}$ is obtained using the \emph{ffs} instruction~\cite{PROGRAMGUIDE} (line 7). This index is the index of the cache block whose metadata matched the request's tag. Finally, the LRU counters are updated (lines 8-12).

\begin{algorithm}[h]
\scriptsize
\setstretch{1.1}
\caption{Extended LLC Tag Lookup -- Register File}
\label{algo:hk1}
\centering
\textbf{Input: Extended LLC Request's Tag ($R_{aux_0}$)} \\
\textbf{Output: HIT:bool, and if HIT=True, BLOCK\_INDEX:int ($R_{aux_3}$)}
\begin{algorithmic}[1]
\Procedure{Tag Lookup}{}\algcomment{executed by an \hkernel warp of 32 threads}
        \State $R_{aux_1} \gets Valid(R_{M})$ \algcomment{ensure the block is valid}
        \State $R_{aux_1}  \gets R_{aux_1}\ \&\&\ \left(R_{aux_0} == Tag(R_{M})\right)$ \algcomment{match request tag to metadata}
        \State $R_{aux_2} \gets \_\_ballot\_sync(0xffffffff, R_{aux_1})$ \algcomment{share $R_{aux_1}$ between all threads as a 32-bit vector}
        \If{($R_{aux_2}$)} \algcomment{one of the bits is non-zero because there was a hit} 
                 \State $R_{aux_3} \gets \_\_ffs(R_{aux_2}) - 1$ \algcomment{get the 0-based index of the non-zero bit}
                 \State HIT $\gets$ True
                 \State BLOCK\_INDEX $\gets$ $R_{aux_3}$
                 \If{($thread\_idx == R_{aux_3}$)} \algcomment{reset the LRU counter of the hit block} 
                    \State $LRU\_Counter(R_{M}) \gets 0xfff$
                    \Else \algcomment{decrement the LRU counters of all other blocks}
                    \State $ LRU\_Counter(R_{M}) \gets LRU\_Counter(R_{M}) - 1$
                    \EndIf
                \Else  
                \State HIT $\gets$ False
                
            \EndIf
\EndProcedure

\end{algorithmic}
\end{algorithm}

\noindent\textbf{Handling Extended LLC Hits.} 
After detecting an extended LLC hit, the corresponding warp in the \hkernel should move the cache block from the register file to the \emph{read data buffer} in the \titleShort{} controller~(\S\ref{sec:warp-status}). In this section, we explain how the warp in the \hkernel reads the register that contains the requested cache block.

After the \emph{tag lookup} procedure, the corresponding \hkernel warp has the index of the matching cache block available (e.g., in $R_{aux_3}$). To retrieve the matching cache block's data, the \hkernel warp should access the register whose index equals the value of $R_{aux_3}$. For example, if $R_{aux_3}$ is equal to 5, $R_{5}$ should be accessed. This register file access is \emph{indirect}, i.e., it requires reading from a register whose index is determined by accessing the value in another register. An indirect register file access is not straightforward, because many existing GPU ISAs (e.g.,~\cite{PROGRAMGUIDE}) only provide instructions for accessing the register file with an immediate (constant) index.

To enable indirect register accesses, we define a procedure called \emph{Indirect-MOV}. The key mechanism is to implement a \emph{switch-case} structure in the procedure. The procedure 1) allocates each case to access a specific register index, and 2) selects the corresponding case using the target register index (e.g., the value of $R_{aux_3}$). Algorithm~\ref{algo:move} illustrates how we implement the \emph{Indirect-MOV} procedure using the instructions already present in an existing GPU ISA~\cite{PROGRAMGUIDE}.\footnote{We optimize \emph{Indirect-MOV} by adding a new instruction to the ISA in \S\ref{sec:compressor}.} The procedure uses the \emph{brx.idx} instruction~\cite{PROGRAMGUIDE}. 
This instruction is a branch instruction that gets a list of branch targets (i.e., T$_{List}$) and an index as input. The control flow jumps to the branch targets at the specified index. The \emph{Indirect-MOV} procedure uses the the target LLC block index (e.g., the value of $R_{aux_3}$) as the input to the \emph{brx.idx} instruction.   
The procedure defines 32 branch targets ($L_{0}$-$L_{31}$), each is allocated to access the corresponding data-array register. For example, the branch targets \emph{L0} and \emph{L31} are to access the cache blocks in registers $R_{0}$ and $R_{31}$, respectively.

\begin{algorithm}[h]
\scriptsize
\setstretch{1}
\caption{Indirect-MOV Algorithm}
\label{algo:move}
\centering
\textbf{Input: BLOCK\_INDEX:int ($R_{aux_3}$)} \\
\textbf{Output: Requested Extended LLC Block ($R_{aux_0}$)}

\begin{algorithmic}[1]

\Procedure{Indirect-MOV}{}
   \algcomment{The goal is to implement register indirect access, reading from a register
whose index is determined by accessing the value in another
register. This procedure is critical for accessing data-array registers in the extended LLC kernel.} \vspace{1mm}
   \State $T_{list}:\,\, .Branch \,\, Targets\,\, L_0, L_1, L_2, …, L_{31};$ \algcomment{Define 32 branch targets, each is allocated to access a specific register index.} \vspace{1mm}
    \State $@p\,\, brx.idx \,\,\, R_{aux_3}, T_{list};  $ \algcomment{Branch to label $L_{i}$ specified by the target LLC block index $i$=$R_{aux_3}$} \vspace{1mm}
    \State $L_0:$
    \State $\, \, \, \, MOV \,\,\, R_0, R_{aux0}$ \algcomment{Access data-array register $R_0$ if target LLC block index is 0}
    \State $\, \, \, \, return$ \vspace{1mm}
    \State $L_1:$
    \State $\, \, \, \, MOV \,\,\, R_1, R_{aux0}$ \algcomment{Access data-array register $R_1$ if target LLC block index is 1}
    \State $\, \, \, \, return$ \vspace{1mm}

    \State \,\,\,...
    \State $L_{31}:$
    \State $\, \, \, \, MOV \,\,\, R_{31},R_{aux0}$\algcomment{Access\,data-array\,register\,$R_{31}$\,if\,target\,LLC block\,index\,is\,31}
    \State $\, \, \, \, return$ \vspace{1mm}

\EndProcedure
\end{algorithmic}
\end{algorithm}

\noindent\textbf{Handling Extended LLC Misses.} 
The warp in the \hkernel that services the extended LLC request handles an extended LLC miss in four steps. First, the warp accesses main memory to bring the requested cache block. Note that main memory accesses from the \hkernel bypass the conventional LLC. Second, to free space in the extended LLC for the to-be-inserted block, the warp selects the victim extended LLC block based on the \emph{LRU} replacement policy. To this end, the warp determines which extended LLC block has the lowest LRU counter. Third, the warp checks the dirty bit of the victim extended LLC block and writes it back to the main memory if it is dirty. Fourth, the warp writes the new extended LLC block into the register file using the \emph{Indirect-MOV} procedure~(Algorithm~\ref{algo:move}).

\subsubsection{Extended LLC via Unified L1/Shared-memory}
\label{sec:SC-L1}
We explain how the \hkernel uses the L1 cache and shared memory as the extended LLC.\\ %
\textbf{L1 cache.} 
\titleShort{} assigns a portion of the extended LLC to the L1 cache of each GPU core that is in \cachemode. When a warp executing the \hkernel receives a request that should be serviced in the L1 cache, the warp simply  forwards the request to the L1 cache by executing GPU load and store instructions. If the request hits in the L1 cache, the responsible warp in the \hkernel responds to the \titleShort{} controller through the extended LLC query logic unit. Otherwise, the L1 cache accesses the main memory to service the miss request. Note that the \titleShort{} controller ensures that an L1 cache miss from a GPU core that is in \cachemode bypasses the conventional LLC and directly accesses main memory.

\noindent\textbf{Shared Memory.} 
\titleShort{} assigns a portion of the extended LLC to the shared memory. Since shared memory does not have a hardware unit for storing the tags (unlike the L1 cache), the \hkernel stores the tags of the extended LLC blocks assigned to the shared memory inside the register file instead. The advantage of this approach is that a register file access is faster than shared memory access, which accelerates the tag lookup procedure. The tag lookup procedure is similar to the procedure shown in Algorithm~\ref{algo:hk1}.
To access an extended LLC block, the \hkernel calculates the address of the block's data in shared memory based on the extended LLC set number and the cache block index from the tag lookup procedure.

\subsubsection{Supporting Atomic Instructions in the Extended LLC}\label{sec:supporting_atomics} Modern GPUs %
execute global memory atomic operations %
via single SASS instructions %
that run on atomic units in the conventional LLCs~\cite{jia2019dissecting,7476485}, which is critical for performance of GPU applications with inter-CTA synchronization.  
\titleShort{} supports global memory atomic operations in the extended LLC, as we explain next.  
First, the warp handling an extended LLC request performs atomic operations using the functional units inside the SMs in \cachemode, no matter which on-chip memory inside an SM in \cachemode has the corresponding block. 
Second, the \titleShort{} controller guarantees the atomicity in the extended LLC since several threads cannot access the same extended LLC block at the same time. This is because: (1) each cache block in the %
extended LLC %
is assigned to exactly one warp in the \hkernel, %
 and (2) each warp completes one extended LLC request %
before starting %
to service another request. %

\subsection{Optimizing \titleShort}
\label{sec:compressor}
We describe two optimization techniques on top of the basic \titleShort design we introduced in \S\ref{sec:Mech}-\ref{sec:soft-cache}. First, the flexibility of the \hkernel enables low-cost implementation of cache optimization techniques, such as cache compression, resizing cache blocks, and online modification of the replacement policy. As a case study, we discuss how to use cache compression in \titleShort (\S\ref{sec:cache-comp}).  
Second, the extended LLC needs the \emph{Indirect-MOV} procedure that we implement   %
using the \emph{brx.idx} instruction in basic \titleShort %
(see Algorithm~\ref{algo:move}). However, having architectural support for this operation can accelerate the data array access. We provide new architectural support for the \emph{Indirect-MOV} instruction in \S\ref{sec:newMove}.

\subsubsection{Cache Compression}
\label{sec:cache-comp}
Queries to, insertions to, and evictions from the register file and shared memory partitions of the extended LLC \emph{always} go through the \hkernel\footnote{Note that insertions to and evictions from the L1 partition of the extended LLC do \emph{not} go through the \hkernel, because the L1 cache handles them in hardware.}. Hence, the \hkernel can manipulate them in a way that is transparent to the rest of the system. Our goal is to leverage this opportunity for increased extended LLC capacity. To this end, we propose a cache compression scheme on top of \titleShort{}. The key idea is to use the \hkernel to store \emph{compressed} (where possible) versions of the extended LLC blocks, thereby increasing the number of blocks in each extended LLC set, and thus the effective capacity of the extended LLC. The \hkernel compresses any inserted extended LLC block, stores the compressed version in the register file or shared memory, and serves requests by decompressing the blocks upon a hit. In this section, we describe our mechanism in detail.

\pgfkeys{/csteps/inner color=white}
\pgfkeys{/csteps/outer color=black}
\pgfkeys{/csteps/fill color=black}

\begin{figure}[h]
\begin{center}
  \includegraphics[trim= 0mm 0mm 0mm 0mm,width=1\linewidth]{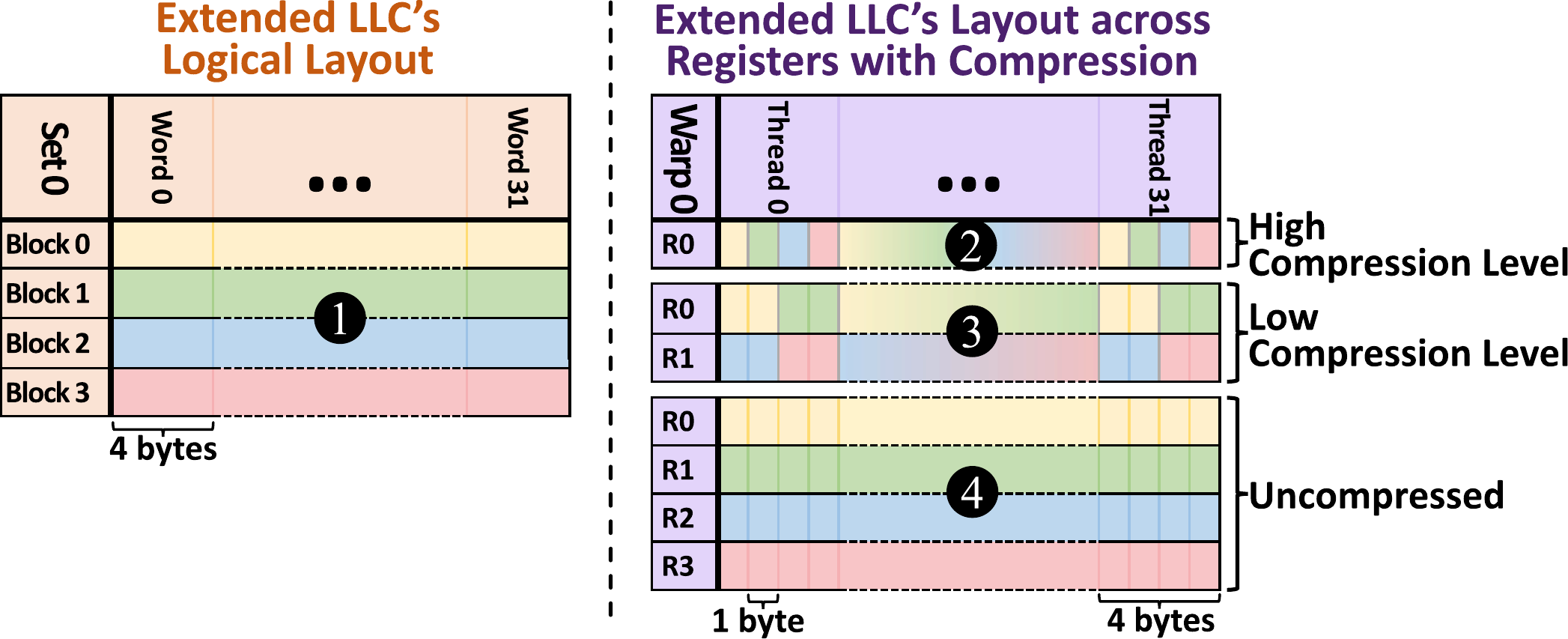}
  \caption{Layout of compressed cache blocks across registers}
  \label{fig:compressor}
\end{center}
\end{figure}

An inserted or updated (i.e., written-to) 128-byte extended LLC block is grouped into one of three \emph{compression levels} that we define as follows: (1)~the \emph{high} compression level includes extended LLC blocks that can be compressed 4-fold into 32 bytes, (2)~the \emph{low} compression level includes extended LLC blocks that can be compressed 2-fold into 64 bytes, and (3)~the \emph{uncompressed} level includes extended LLC blocks that could not be compressed. Figure~\ref{fig:compressor} shows how logical 128-byte extended LLC blocks~\Circled{1} are laid out in the register file, depending on their respective compression level~\Circled{2}\Circled{3}\Circled{4}. Blocks in the \emph{high}~\Circled{2} and \emph{low}~\Circled{3} compression levels are laid out across registers in a strided and interleaved manner with strides 4 and 2, respectively. Blocks in the \emph{uncompressed} level are laid out exactly like they were without our compression scheme~\Circled{4}.
For example, for the \emph{high} compression level, four extended LLC blocks~\Circled{1} are stored in a single warp register of 32$\times$4 bytes~\Circled{2}, such that they occupy the first, second, third and fourth bytes of each thread, respectively.

Since the compression levels of cache blocks cannot be known ahead of time, the number of warp registers allocated to each compression level should be adapted dynamically. Our mechanism initially assigns all registers to the \emph{uncompressed} level. Then, over epochs of \emph{n} cycles (we empirically choose \emph{n}=10,000), the number of cache blocks in the \emph{high}, \emph{low} and \emph{uncompressed} levels are counted. At the end of each epoch, the number of registers assigned to each compression level is updated based on the counter values. %

We employ the Base-Delta-Immediate (BDI) compression algorithm~\cite{pekhimenko2012base} due to its simplicity and good cache compression ratios. The BDI algorithm works as follows: First, the input cache block is divided into segments (e.g., 4 bytes each). Second, one of these segments (e.g., the first) is designated as the \emph{base} segment and copied to the output. Third, only the \emph{deltas} (arithemtic differences) of the remaining segments from the base segment are copied to the output~\cite{pekhimenko2012base}. The achieved compression ratio depends on how large the deltas are. For example, if all segments are very similar to the base segment, the deltas are small and can be stored in only a few bits for each delta.
We consider 4-byte segments in our implementation and store the base segments of compressed blocks in auxiliary registers.

\subsubsection{Indirect-MOV Instruction}
\label{sec:newMove}

The Indirect-MOV procedure~(\S\ref{sec:SC-RF}) is required for indirectly addressing registers in the \hkernel. Our software implementation of the Indirect-MOV procedure (Algorithm~\ref{algo:move}) is very portable and flexible because it uses only instructions from NVIDIA's existing PTX ISA~\cite{PROGRAMGUIDE}. However, it is inefficient and slow for two reasons. First, it executes three instructions (i.e., \emph{brx.idx}, \emph{MOV}, and \emph{return}) to perform a single indirect register access. Second, two of these instructions (i.e., \emph{brx.idx} and \emph{return}) are branches, which cause irregular control flow.

To improve the efficiency of the Indirect-MOV procedure, we introduce a new instruction in the GPU ISA that can perform Indirect-MOV natively via minor modifications to existing hardware. Like the software implementation of Indirect-MOV, the new Indirect-MOV instruction conceptually (1)~accesses the register file to read the source register $R_{src}$, (2)~re-accesses the register file to read the indirectly addressed register $R[R_{src}]$, and (3)~moves the value from $R[R_{src}]$ to the destination register $R_{dest}$.

\begin{figure}[h]
\begin{center}
  \includegraphics[width=1\columnwidth]{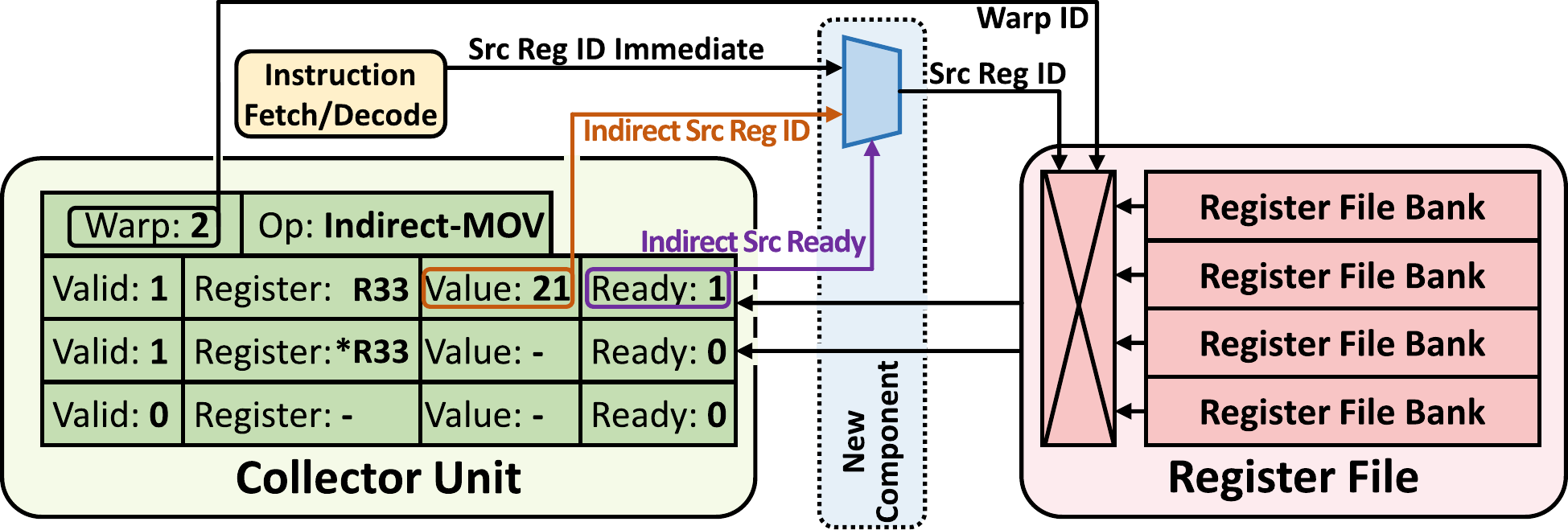}
 \caption{Native hardware implementation of Indirect-MOV} %
  \label{fig:RF-NEWMOVE}
\end{center}
\end{figure}

To support this instruction in GPU hardware, we modify the register file architecture. The key idea is to support two \emph{sequential} register file reads for the Indirect-MOV instruction. We slightly modify the operand collectors in the register file to support these sequential reads. The operand collector first accesses the register file using the register number specified in the instruction to read a 1024-bit warp register. Then, the operand collector uses the first eight least significant bits of the read register value as the next register number. The operand collector then re-accesses the register file with this new register number. The read value is then written to the destination register using the regular MOV instruction's data path in the pipeline.
Figure~\ref{fig:RF-NEWMOVE} illustrates our changes to the baseline operand collector. We add a single multiplexer per operand collector to select between the two different sources of the register number: (1)~the immediate source register number from the instruction, (2)~the value loaded from a register. The multiplexer is controlled by the ready bit of the register number loaded from a register, i.e., the indirect register number is used as soon as it is available. %

 \vspace{-0.8ex}\section{Characterization of the Extended LLC Kernel}
\label{sec:SC-study}
We implement the extended LLC kernel, as described in \S\ref{sec:soft-cache}, and evaluate it on a real GPU. 
Our goal is to obtain relevant metrics that characterize the extended LLC kernel and the different implementation alternatives (i.e., combinations of register file, L1, and shared memory). 
We evaluate four relevant metrics for the extended LLC: (1) storage capacity, (2) access latency, (3) access bandwidth, and (4) energy per byte. 
We use these metrics to (i) determine the implementation alternative that provides the best tradeoff, and (ii) properly configure our simulation of the extended LLC in our cycle-level GPU simulator (see \S\ref{sec:Method}), to evaluate the effectiveness of \titleShort{} at boosting GPU performance of real-world memory-bound applications (\S\ref{sec:Eval}).

\noindent\textbf{Methodology.} 
We implement and evaluate the extended LLC kernel on a real NVIDIA RTX 3080 GPU~\cite{3080-white}. 
Our implementation faithfully follows the description in \S\ref{sec:soft-cache}. 
However, since there is no actual \titleShort{} controller (\S\ref{sec:morpheus}) in a real state-of-the-art GPU, we emulate the warp status table (\S\ref{sec:warp-status}) by placing a similar data structure, which contains the addresses of the extended LLC to access in our evaluation, in the conventional LLC. The latency of an access to this emulated warp status table is similar to an access to the warp status table in the \titleShort{} controller, since the \titleShort{} controller sits inside the LLC partition (see Figure~\ref{fig:proposed-GPU}).\footnote{We ensure that (1) the data structure that emulates the warp status table resides in the conventional LLC (and not L1 cache) using the \emph{ld.global.cg} instruction~\cite{cuda-toolkit}, and (2) the data structure fits completely in the conventional LLC so that all accesses (after the initialization phase of the extended LLC kernel) to this data structure hit in the conventional LLC.
}  

We implement three variants of the extended LLC kernel: (1) extended LLC via register file, (2) extended LLC via L1, and (3) extended LLC via shared memory. 
For each of them, we experiment with different numbers of warps of the extended LLC kernel on a single GPU core. Each warp is in charge of one set of the extended LLC. Thus, the larger the number of warps, the smaller the extended LLC sets are (because of the fixed size of memory storage). 
Each extended LLC variant and number of warps offers a different tradeoff in terms of capacity, latency, bandwidth, and energy/byte.

To measure the extended LLC capacity, we calculate the available space for the extended LLC data array per GPU core in \cachemode. This depends on the size of the actual storage (i.e., register file, shared memory, L1) and the space needed for \emph{auxiliary} purposes (e.g., the execution context of \hkernel).  %
To measure the extended LLC access latency, we use the \emph{Nsight} profiler tool~\cite{iyer2016gpu} and the \emph{cudaEventElapsedTime} API~\cite{cuda-toolkit}. 
To measure the extended LLC access bandwidth, we first measure the number of \emph{accesses per second} by dividing the total number of extended LLC accesses (100 million accesses in our experiments) by the total time takes to service all extended LLC accesses. Second, we multiply the resulting accesses per second by 128 (the extended LLC block size in bytes) to calculate the extended LLC bandwidth in bytes per second (B/s). 
To measure the extended LLC energy per byte, we first measure the average GPU power consumption using \emph{nvidia-smi}~\cite{PROGRAMGUIDE} while  servicing 100 million accesses to the extended LLC. Second, we calculate the total energy consumption by multiplying the measured power consumption with the total time it takes to service these extended LLC requests. Third, we divide the total energy consumption by the number of extended LLC accesses to obtain the energy per access. Fourth, we divide the energy per access by 128 (the extended LLC block size in bytes) to calculate the extended LLC energy per byte (J/B).

\begin{figure}[h]
\centering
\includegraphics[trim = 0mm 0mm 0mm 0mm, clip, width=1\linewidth, right]{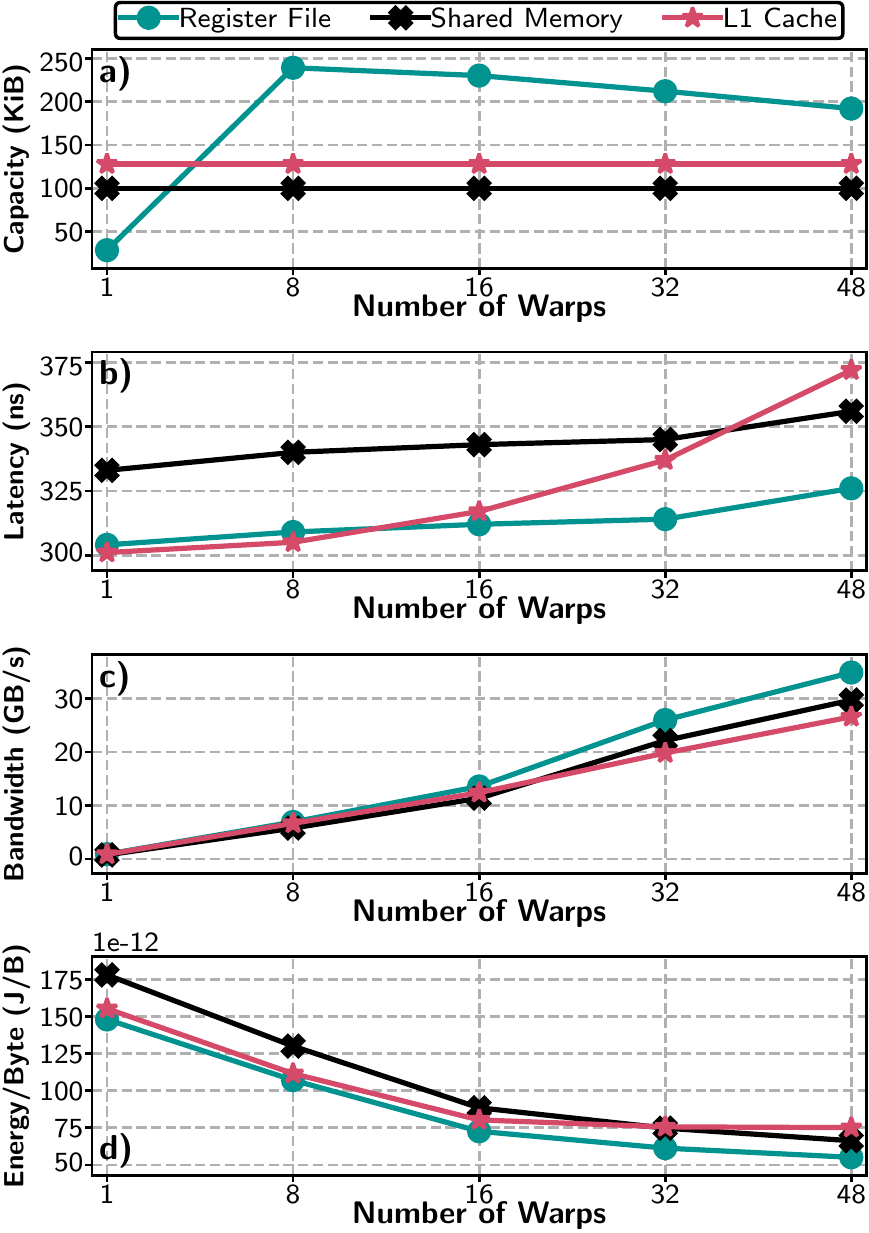}
\caption{Characterization of the extended LLC using a real GPU~\cite{3080-white}. a) extended LLC capacity, b) extended LLC access latency, c) extended LLC access bandwidth, and d) extended LLC energy per byte.} %
\label{fig:software-LLC-study}
\end{figure}

\noindent\textbf{Extended LLC Capacity.} Figure~\ref{fig:software-LLC-study}(a) reports the extended LLC capacity per GPU core in \cachemode for different implementations using various numbers of warps, i.e., 1, 8, 16, 32, and 48.  %
We make four key observations. First, the extended LLC capacity is substantial per GPU core in \cachemode. For example, when using 8 warps for the extended LLC via register file (providing 239 KiB capacity) and 8 warps for the extended LLC via L1 (providing 128 KiB capacity), the extended LLC capacity is 367 KiB per GPU core in cache mode.\footnote{Note that using shared memory would not further increase the extended LLC capacity in this case. This is because the L1 and shared memory are unified in modern NVIDIA GPUs, i.e., the sum of L1 and shared memory space in a core is at most their total unified storage capacity (e.g., 128 KiB in an NVIDIA RTX 3080~\cite{3080-white}).}
Second, the capacity of the extended LLC via register file varies with the number of warps. This is due to two main reasons. First, using fewer than eight warps, the \hkernel cannot utilize the total register file capacity since the extended LLC capacity is limited to the maximum number of registers per thread (i.e., 256). Third, using eight warps results in the maximum extended LLC capacity via register file (i.e., 239 KiB). Using more than eight warps leads to smaller extended LLCs via register file due to allocating a higher number of registers for \emph{auxiliary} purposes (e.g., the execution context of the \hkernel warps). 
Fourth, the capacity of the extended LLC via L1 and via shared memory does \emph{not} change with the number of warps. This is because the \hkernel allocates the whole space of each of these two memories (L1 and shared memory) to the extended LLC data array no matter how many warps the kernel uses. %

\noindent\textbf{Extended LLC Access Latency and Bandwidth.} Figures \ref{fig:software-LLC-study}(b) and~\ref{fig:software-LLC-study}(c) report the extended LLC access latency and bandwidth, respectively, for different implementations using various numbers of warps, i.e., 1, 8, 16, 32, and 48. We make five key observations. First, the extended LLC access latency ($\geq$300 ns) is almost two times longer than the conventional LLC access latency ($\sim$160 ns~\cite{jia2019dissecting,yan2020optimizing}). The longer access latency of the extended LLC compared to the conventional LLC is mainly because of the round trip interconnect latency from the \titleShort{} controller to the GPU core in \cachemode and from the GPU core in \cachemode to the \titleShort{} controller. However, the extended LLC access latency is still approximately 2$\times$ faster than accessing off-chip memory ($\sim$600ns~\cite{jia2019dissecting,yan2020optimizing}). Second, the extended LLC access bandwidth per GPU core in \cachemode is 37 GB/s using the register file implementation and 48 warps. The bandwidth of each conventional LLC partition is around 300 GB/s~\cite{jia2019dissecting,yan2020optimizing}, and thus eight GPU cores in \cachemode can collectively provide the same bandwidth as one conventional LLC partition. 
Third, increasing the number of warps of the \hkernel in all three implementations results in higher extended LLC bandwidth at the cost of longer access latency. Increased access latency with more warps is mainly due to the fact that the corresponding warp servicing a request to the extended LLC needs to wait until its scheduling slot, and the waiting time becomes longer when using more warps. Fourth, the extended LLC via register file has both lower latency and higher bandwidth 
for almost all warp counts, compared to the extended LLC via shared memory and via L1.  %
This is because the register file has a lower access latency and a higher access bandwidth compared to shared memory and L1.\footnote{The access latency of the register file, shared memory, and L1 are 2, 25, and 34 ns, respectively~\cite{khairy2020accel}. The access bandwidth of the register file, shared memory, and L1 are 1 TB/s, 170 GB/s, and 170 GB/s, respectively~\cite{khairy2020accel,jia2019dissecting,yan2020optimizing}.} Fifth, the bandwidth of the extended LLC in the best case (i.e., the extended LLC via register file using 48 warps) is still less than 40 GB/s, which is one order of magnitude lower than the bandwidth of the register file (1 TB/s). This is because 
the interconnection network connecting LLC partitions to GPU cores significantly bottleneck the bandwidth of the extended LLC via register file (and similarly via shared memory and via L1). 

To further analyze the effect of the interconnection network, we \emph{ideally exclude} the interconnection network from the extended LLC accesses. To this end, we (1)~assume the address of the extended LLC request is already ready in an \emph{auxiliary} register of the SM operating in \cachemode and (2)~let the SM operating in \cachemode discard the response instead of sending it over the interconnection network. We observe that the access bandwidth of the extended LLC via register file, shared memory, and L1 using 48 warps becomes 290 GB/s, 106 GB/s, and 97 GB/s, which is 7.8$\times$, 3.4$\times$, and 3.5$\times$ higher than the \emph{non-ideal} versions, respectively. Hence, a better interconnect design could significantly improve the performance of the extended LLC.

\noindent\textbf{Extended LLC Energy per Byte.} Figure \ref{fig:software-LLC-study}(d) reports energy per byte results for different implementations using various numbers of warps, i.e., 1, 8, 16, 32, and 48. Note that the energy per byte results take into account the energy consumed by all the components included in the extended LLC accesses, i.e., GPU cores in \cachemode (executing the \hkernel), interconnect, and the conventional LLC banks. We make three key observations. First, The extended LLC energy per byte in the best case (extended LLC via register file and using 48 warps) is 53 pJ which is approximately 5.3$\times$ the energy per byte of the conventional LLC ($\sim$10 pJ~\cite{kandiah2021accelwattch}). Despite its high energy consumption compared to the conventional LLC, the extended LLC can reduce GPU energy consumption by reducing the number of energy-hungry off-chip memory accesses. Second, increasing the number of warps in the \hkernel significantly reduces energy per byte for all implementations. This is because using more warps increases the extended LLC's throughput, without significantly increasing power consumption, leading to lower energy per byte. Third, the extended LLC via register file has a lower energy per byte for all warp counts compared to the extended LLC via L1 and via shared memory. This is mainly due to the fact that a register file access consumes less energy compared to an access to the L1 or shared memory~\cite{kandiah2021accelwattch}. 

\noindent\textbf{Combining different extended LLC versions.} Our characterization of the extended LLC shows that the extended LLC via register file outperforms other implementations in terms of capacity, access latency, access bandwidth, and energy per byte. However, to utilize all on-chip memories of a GPU core in \cachemode and thus enable larger extended LLC capacities, the \hkernel aims to combine the implementation via register file with the implementation via L1 and/or via shared memory. To this end, the \hkernel allocates a number of warps to the implementation via register file, and the remaining number of warps to the implementations via L1 and shared memory. We only consider the combination of the extended LLC via register file and via L1 since L1 and shared memory are unified in our evaluated GPU system (i.e., combining the extended LLC via L1 and via shared memory does \emph{not} provide a larger extended LLC capacity).

We use our characterization results to optimize the number of warps the \hkernel should allocate to each of the two \hkernel implementations (via register file and via L1) to design a high-capacity and high-performance extended LLC. (1) The \hkernel should allocate more than 8 warps to the implementation via register file to better utilize the register file capacity (Figure~\ref{fig:software-LLC-study}(a)). (2) The \hkernel should allocate fewer than 48 warps to the implementation via register file to be able to allocate a number of warps to the implementation via L1. (3) Among the remaining options (8, 16, or 32 warps), using 32 warps in the \hkernel via register file results in higher access bandwidth (Figure~\ref{fig:software-LLC-study}(c)) and lower energy per byte (Figure~\ref{fig:software-LLC-study}(d)), while the access latency is almost equal to the case of using 8 warps (Figure~\ref{fig:software-LLC-study}(b)). Hence, the \hkernel combines the extended LLC via register file and via L1 by allocating 32 and 16 warps to each of these two implementations, respectively. Using our characterization, we observe that for each GPU core operating in \cachemode, the extended LLC via \emph{register file+L1} has 328 KiB capacity, 185ns average access latency, 34GB/s average access bandwidth, and 61pJ average energy per byte. We use this extended LLC implementation to evaluate the effectiveness of \titleShort{} to boost the performance of memory-bound applications.

\section{Methodology}
\label{sec:Method}
\noindent\textbf{Simulation methodology.} We evaluate \titleShort{} using the AccelSim~\cite{khairy2020accel} cycle-level simulator. We model our baseline after the NVIDIA Ampere 3080 GPU~\cite{3080-white}. Table~\ref{tab:software-cache-analysis} shows the simulation parameters modeling our baseline. 
To evaluate energy consumption, we use AccelWattch~\cite{kandiah2021accelwattch} embedded in AccelSim as the state-of-the-art GPU energy model.

\begin{table}[h!]
  \centering
  \caption{Baseline GPU configuration} %
  \small
  \vspace{-1mm}
  \begin{tabular}{|l|l|}
    \hline
    \textbf{Parameter} & \textbf{Value}\\
    \hline
    \hline
    Number of SMs & 68\\
    \hline
    Scheduler & Two-Level~\cite{narasiman2011improving,gebhart2011energy} \\ \hline
    GPU Memory Interface & 320-bit GDDR6X~\cite{GDDR6X}\\
    \hline
    GPU Memory Capacity & 10 GiB\\
    \hline
    Conventional LLC Capacity & 5 MiB\\
    \hline
    L1/Shared-Memory Capacity & 128 KiB per SM\\
    \hline
    Register File Capacity & 256 KB per SM\\
    \hline
  \end{tabular}
  \label{tab:software-cache-analysis}
\end{table}

\noindent\textbf{Applications.} 
We randomly choose 14 \emph{memory-bound} and 3 \emph{compute-bound} applications from four benchmark suites, Rodinia~\cite{che2009rodinia}, Parboil~\cite{stratton2012parboil}, Pannotia~\cite{che2013pannotia} and ISPASS~\cite{bakhoda2009analyzing}. Our methodology to categorize applications into memory-bound and compute-bound groups is based on our experiment in \S\ref{sec:motiv}. The performance of the compute-bound applications to increases (linearly) with more GPU cores. In contrast, the performance of the memory-bound applications either saturates or decreases sharply after a certain number of GPU cores. Table~\ref{tab:applications} shows the applications we choose, their names, and their types (memory-bound or compute-bound). %
We run each application either entirely, or until the application reaches two billion executed instructions. 

\begin{table}[h!]
  \centering
  \caption{Evaluated Applications}
 
  \resizebox{1\columnwidth}{!}{%
  \begin{tabular}{|l|l|l|l|}
    \hline
    \textbf{Application} & \textbf{Name}& \textbf{Type}\\
    \hline
    \hline
    Breadth-First Search\cite{stratton2012parboil} & p-bfs & Memory-bound\\
    \hline
      Computational fluid dynamics\cite{che2009rodinia} & cfd & Memory-bound\\
    \hline
     Discrete Wavelet Transform (2D)\cite{che2009rodinia} & dwt2d & Memory-bound\\
    \hline
     Stencil\cite{stratton2012parboil} & stencil & Memory-bound\\
    \hline 
    Breadth-First Search\cite{che2009rodinia} & r-bfs & Memory-bound\\
    \hline
    Back Propagation\cite{che2009rodinia} & bprob & Memory-bound\\
    \hline
    sgemm\cite{stratton2012parboil} & sgem & Memory-bound\\
    \hline
     Needleman-Wunsch\cite{che2009rodinia} & nw & Memory-bound\\
    \hline
     Page Rank\cite{che2013pannotia} & page-r & Memory-bound\\
    \hline
      K-means\cite{che2009rodinia} & kmeans & Memory-bound\\
    \hline
    Histogram\cite{stratton2012parboil} & histo & Memory-bound\\
    \hline
     Magnteic Resonance Imaging-Gridding\cite{stratton2012parboil} & mri-gri & Memory-bound\\
    \hline
    Sparse-Matrix Dense-Vector Multiplication\cite{stratton2012parboil} & spmv & Memory-bound\\
    \hline
     Lattice-Boltzmann\cite{stratton2012parboil} & lbm & Memory-bound\\
    \hline
     LIBOR Monte Carlo\cite{bakhoda2009analyzing} & lib &Compute-bound\\
    \hline
     HotSpot\cite{che2009rodinia} & hotsp & Compute-bound\\
    \hline
    Magnetic Resonance Imaging - Q\cite{stratton2012parboil} & mri-q & Compute-bound\\
    \hline
    
    \hline
  \end{tabular}
  \label{tab:applications}
  }
\end{table}

\noindent\textbf{Evaluated Systems.} We evaluate six systems. (1) \emph{BL}: baseline system that models a GPU architecture with the parameters reported in Table~\ref{tab:software-cache-analysis}. \emph{BL} employs all available GPU cores (i.e., 68) for application execution. To provide a fair comparison, we add the extra on-chip storage in \titleShort{}, the 16-KiB Bloom filters (\S\ref{sec:misspenalty}) and the 5-KiB extended LLC query logic unit (\S\ref{sec:warp-status}) per LLC partition (overall; 21~KiB$\times$\#partitions = 210~KiB), to the conventional LLC capacity. (2) \emph{IBL}: improved baseline system %
where we use the number of GPU cores  that provides the maximum performance for each application and power-gate the remaining cores. Table~\ref{tab:cache-sm-count} (second row) reports the number of GPU cores we use for each application in \emph{IBL}.  %
\begin{table*}[h]
  \centering
  \scriptsize
  \caption{Number of GPU cores executing application threads for different evaluated systems (\#available GPU cores is 68).}
  \resizebox{\textwidth}{!}{%
  \setlength\tabcolsep{1.5pt}
  \begin{tabular}{|l||r|r|r|r|r|r|r|r|r|r|r|r|r|r|r|r|r|}
    \hline
    \textbf{Application} & p-bfs & cfd & dwt2d & stencil & r-bfs & bprob & sgem & nw & page-r & kmeans & histo & mri-gri & spmv & lbm & lib & hotsp & mri-q\\
    \hline\hline
    \textbf{IBL} & 68 & 68 & 68 & 68 & 68 & 68 & 68 & 68 & 68 & 24 & 53 & 34 & 42 & 34 & 68 & 68& 68\\
    \hline
    \textbf{\titleShort{}-Basic} & 32 & 42 & 42 & 50 & 34 & 39 & 48 & 18 & 42 & 37 & 47 & 36 & 44 & 32 & 68 & 68& 68\\
    \hline
    \textbf{\titleShort{}-ALL} & 40 & 55 & 54 & 56 & 37 & 41 & 54 & 26 & 46 & 47 & 52 & 43 & 47 & 36 & 68 & 68 & 68 \\
    \hline
  \end{tabular}
  \label{tab:cache-sm-count}
  }
\end{table*}
(3) \emph{IBL-4$\times$-LLC}: \emph{IBL} with 4$\times$ the LLC size. We increase the total LLC capacity by increasing the number of the LLC banks (without adding any latency and power impact). (4) \emph{Frequency-Boost}: \emph{IBL} with higher frequency memory system components, including the interconnection network, conventional LLC, and off-chip DRAM channels. This system uses the energy saved by power-gated GPU cores in \emph{IBL} to increase the clock frequency of the aforementioned components in the GPU memory system by 10\%-20\% depending on the number of power-gated cores. (5) \emph{Unified-SM-Mem}: \emph{IBL} that has a larger L1 data cache capacity by using methods from prior works on unifying L1 data cache, shared memory, and the register file~\cite{gebhart2012unifying,jing2016cache}. Our baseline architecture (\emph{BL}) already unifies L1 data cache and shared memory. In \emph{Unified-SM-Mem}, we add the amount of unused register file space to the L1 data cache (without additional latency impact). %
(6) Different versions of \titleShort with and without our various optimizations, namely \emph{\titleShort-Basic}, \emph{\titleShort-Indirect-MOV}, \emph{\titleShort-Compression}, and \emph{\titleShort-ALL}. For all \titleShort variants, %
we determine the number of GPU cores in \cachemode that results in the highest performance per application offline.\footnote{This is a static version of \titleShort{} in which we adjust the number of SMs in \cachemode before running the application. If the best configuration per application is not known prior to execution, we could use online profiling techniques similar to prior work~\cite{xu2016warped,chang2016dysel,park2017dynamic} on top of \titleShort{}, to learn the best configuration for a running application and dynamically adjust the number of cores in \cachemode. We leave the use of online profiling techniques with \titleShort to future work.} %
Table~\ref{tab:cache-sm-count} (third and fourth rows) reports the number of GPU cores in \computemode for \emph{\titleShort{}-Basic} and \emph{\titleShort{}-ALL}. We observe that \emph{\titleShort{}-ALL} uses a larger number of GPU cores in \computemode because it employs cache compression (\S\ref{sec:cache-comp}), which enables larger extended LLC capacities per SM in \cachemode, which in turn enables higher performance with more application threads. %

\section{Evaluation}
\label{sec:Eval}
We evaluate the effectiveness of \titleShort{} compared to different baselines. \S\ref{sec:perf} shows the overall effect of \titleShort{} on GPU performance. \S\ref{sec:energy} evaluates the effect of \titleShort{} on GPU energy efficiency. \S\ref{sec:costs} analyzes the effectiveness of our Bloom filter-based hit/miss predictor design. \S\ref{sec:BW-Eval} shows the effect of \titleShort{} on on-chip and off-chip bandwidth utilization. \S\ref{sec:overhead} studies the storage and power overheads of \titleShort{}.
\subsection{Performance Analysis}
\label{sec:perf}
\begin{figure*}[h]
\begin{center}
  \includegraphics[trim= 0mm 0mm 0mm 0mm,width=1\linewidth]{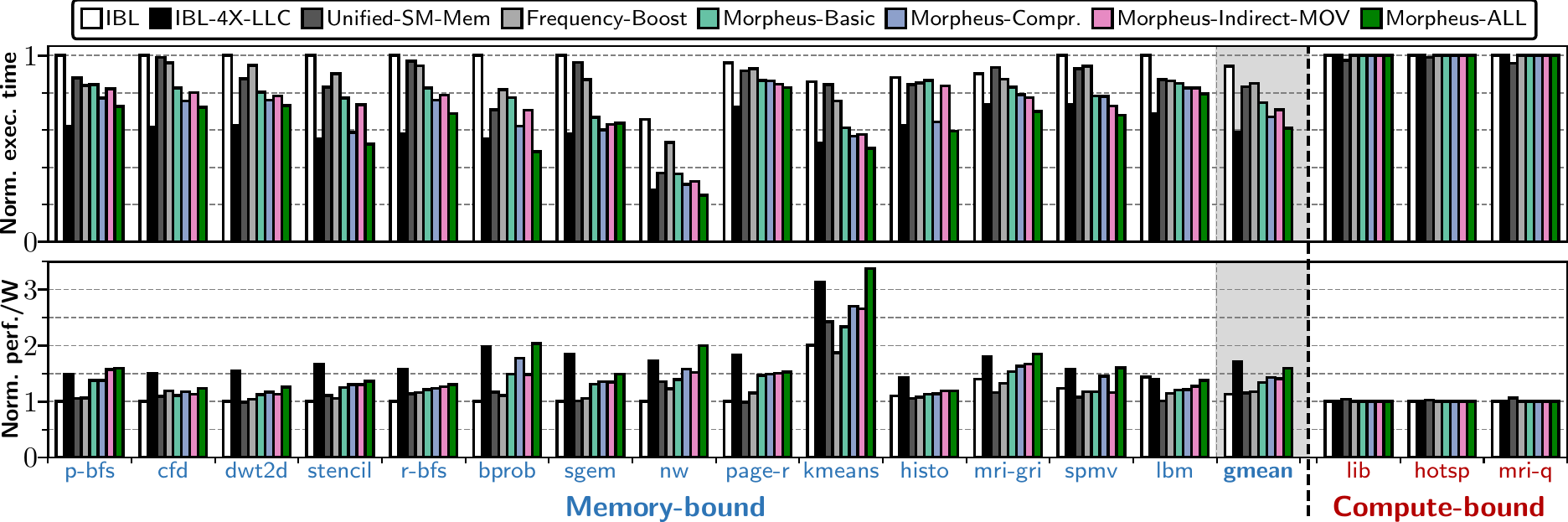}
  \caption{Comparison of eight GPU systems' execution time (top) and performance/watt (bottom) for 14 memory-bound and 3 compute-bound applications, normalized to the baseline system (BL)}
  \label{fig:comp-perf}
\end{center}
\end{figure*}

To study the effectiveness of \titleShort{} at improving GPU performance, we measure application execution times on \emph{nine} different systems, namely \emph{BL}, \emph{IBL}, \emph{IBL-4$\times$-LLC}, \emph{Frequency-Boost}, \emph{Unified-SM-Mem}, \emph{\titleShort{}-Basic}, \emph{\titleShort{}-Compression}, \emph{\titleShort{}-Indirect-MOV}, and \emph{\titleShort{}-ALL} (see \S\ref{sec:Method} for details of each evaluated system). Figure~\ref{fig:comp-perf} (top) shows the results. 
The x-axis shows applications in two groups, memory-bound and compute-bound. The y-axis shows the application execution time (the lower the better) normalized to the baseline (\emph{BL}) system. %

We make five key observations. First, \titleShort (i.e., \emph{\titleShort-ALL}) improves performance greatly with all its optimizations over all real baselines and across all memory-bound applications. Specifically, \titleShort{} significantly improves GPU performance by an average of 27\% over the best real baseline (i.e., \emph{Unified-SM-Mem}) and by 39\%, 32\%, and 29\% over \emph{BL}, \emph{IBL}, and \emph{Frequency-Boost}, respectively. Second, \titleShort performs within 3\% of an ideal baseline with 4× the LLC (i.e., \emph{IBL-4$\times$-LLC}) at only small overhead (see \S\ref{sec:overhead}). Third, \emph{\titleShort-Compression} improves performance by an average of 9\% over \emph{\titleShort-Basic}, by providing a larger extended LLC capacity. Fourth, \emph{\titleShort-Indirect-MOV} improves performance by an average of 4\% over  \emph{\titleShort-Basic}, by providing a lower extended LLC access latency. Fifth, \titleShort{} does \emph{not} affect the performance of compute-bound applications since all GPU cores stay in \computemode for such applications. We conclude that \titleShort{} is highly effective at
improving the performance of memory-bound GPU applications.

\subsection{Energy Efficiency Analysis}
\label{sec:energy}
To study the effect of \titleShort on GPU energy efficiency, we calculate GPU \emph{performance/watt} by dividing the overall GPU IPC by GPU average power consumption for each of the nine systems we evaluate. Figure~\ref{fig:comp-perf} (bottom) reports the results, where the y-axis shows performance/watt (the higher the better). We normalize the results to the performance/watt of the baseline (\emph{BL}) system. 

We make four key observations. First, \titleShort (i.e., \emph{\titleShort-ALL}) improves GPU energy efficiency greatly over all real baselines and across all memory-bound applications. \titleShort{} provides 58\%, 38\%, 35\%, and 33\% better energy efficiency compared to \emph{BL}, \emph{IBL}, \emph{Unified-SM-Mem}, and \emph{Frequency-Boost}, on average, respectively. \titleShort{}' energy efficiency improvement is due to (1) reducing the number of energy-hungry off-chip memory accesses and (2) \titleShort{}' large speedups over all real baselines (\S\ref{sec:perf}). Second, the energy efficiency of the \emph{\titleShort-ALL} system is within 6\% of that of the \emph{IBL-4$\times$-LLC} system, for which we \emph{ideally} assume \emph{no} power and latency impact while using a 4$\times$ larger LLC. Third, \emph{\titleShort-Compression} and \emph{\titleShort-Indirect-MOV} improve the energy efficiency of \emph{\titleShort-Basic} by 8\% and 5\%, on average, respectively. 
Fourth, \titleShort slightly reduces the performance/watt of compute-bound applications (less than 1\%) compared to the baseline (\emph{BL}) system due to the power consumption overhead of the \titleShort{} controller (see \S\ref{sec:overhead} for our overhead analysis). A more optimized \titleShort{} can power-gate the \titleShort{} controller completely for compute-bound applications to avoid the power overhead for such workloads~\cite{hu2004microarchitectural,DBLP:conf/IEEEpact/KayiranJPATKLMD16,DBLP:conf/micro/Abdel-Majeed0A13,sadrosadati2019itap}. We conclude that \titleShort is highly effective at improving the energy efficiency of memory-bound GPU applications.

\subsection{Effect of Hit/Miss Prediction} 
\label{sec:costs}
\begin{figure}[h]
\begin{center}
  \includegraphics[trim= 0mm 0mm 0mm 0mm,width=\linewidth]{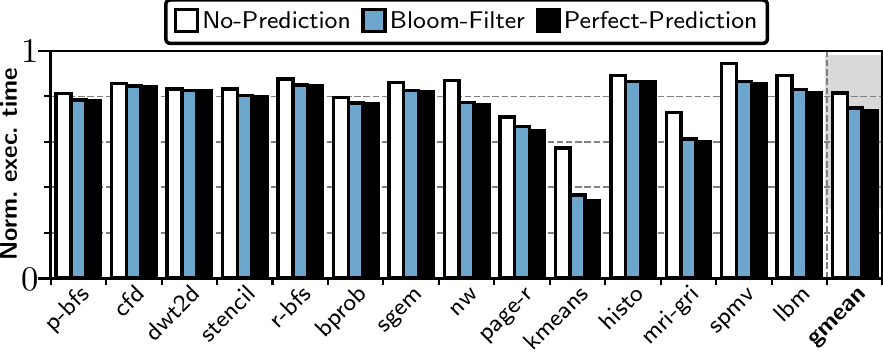}
  \caption{Effect of hit/miss prediction on execution time for 14 memory-bound applications with \emph{\titleShort-Basic}, normalized to the baseline system (BL)}%
  \label{fig:bloom}
\end{center}
\end{figure}

To study the effectiveness of \titleShort{}' hit/miss prediction technique at improving \titleShort{}' performance, we compare the execution time of the 14 memory-bound applications on a \emph{\titleShort-Basic}-enabled GPU with three hit/miss prediction designs: (1)~our \emph{Bloom-Filter} design (\S\ref{sec:misspenalty}), (2)~\emph{No-Prediction}, where we disable the hit/miss prediction technique and immediately forward all extended LLC requests to the extended LLC, and (3) \emph{Perfect-Prediction}, where we assume 100\% accuracy %
for the hit/miss prediction technique.  %
Figure~\ref{fig:bloom} reports the execution time results normalized to the baseline system (\emph{BL}). We make two key observations. First, \emph{No-Prediction} has a 9\% higher execution time compared to the \emph{Bloom-Filter} design, on average. Second, \emph{Bloom-Filter}'s results are within 1\% of \emph{Perfect-Prediction}. We conclude that hit/miss prediction is important for Morpheus’ performance and that our predictor design is very effective. %
\vspace{2em}

\subsection{On-Chip \& Off-Chip Bandwidth Analysis}
\label{sec:BW-Eval}
To further analyze the sources of \titleShort{}' performance %
improvement, we study the effect of \titleShort{} on (1) \emph{LLC throughput}, (2) \emph{interconnect performance}, and (3) \emph{off-chip bandwidth utilization}.

\noindent\textbf{LLC throughput.} In this study, we aim to measure (1) by how much \titleShort{} improves the LLC throughput and (2) determine the reasons for the LLC throughput increase. 
To this end, we measure the LLC (both the conventional LLC and extended LLC) throughput %
for four systems, %
\emph{BL}, \emph{IBL}, \emph{\titleShort{}-ALL}, and  \emph{larger-LLC}. %
The goal of evaluating the \emph{larger-LLC} system is to distinguish between two different benefits of \titleShort{} on LLC (larger LLC capacity and higher number of LLC banks). In \emph{larger-LLC}, we %
increase the conventional LLC size to be exactly the same as the total LLC capacity in \emph{\titleShort{}-ALL} (i.e., conventional LLC + extended LLC) without increasing the number of conventional LLC banks. This is to isolate the effect of an increased LLC capacity (relative to \emph{BL}) from the effect of an increased number of banks.  %
The LLC capacity of \emph{larger-LLC} varies between applications, as it depends on the number of cores operating in \cachemode in \emph{\titleShort{}-ALL}, which is determined on a per application basis.

We make two key observations. First,   %
\emph{\titleShort{}-ALL} improves the LLC throughput by an average of 75\% and 68\% (up to 374\% and 236\%) compared to \emph{BL} and \emph{IBL}, respectively. Second, the \emph{larger-LLC} system improves LLC throughput by an average of 42\% compared to \emph{BL}. We conclude that \emph{\titleShort{}-ALL}'s higher throughput comes from \emph{both} (1)~increasing the LLC capacity and (2)~increasing the number of banks.

\noindent\textbf{Interconnect Performance.} \titleShort{} increases the load on the interconnection network by servicing the extended LLC requests through it. We study the performance of the GPU interconnection network in \titleShort{}.
We measure the overall network injection rate, network throughput, and average interconnect latency for two designs, %
\emph{BL} and \emph{\titleShort{}-ALL}. We make three key observations. First, \emph{\titleShort{}-ALL} increases the load of the on-chip network by 97\% compared to \emph{BL}, on average. Second,  %
both network injection rate and throughput %
increase in \emph{\titleShort{}-ALL} %
by the same amounts, showing that the GPU interconnection network does \emph{not} saturate due to handling more traffic. Third, the higher load causes 7\% longer average network latency compared to \emph{BL}. We observe \emph{no} overall application performance loss due to the longer average network latency.%

\noindent\textbf{Off-chip Bandwidth Utilization.} To study the effect of \titleShort{} on off-chip bandwidth utilization, we measure the off-chip bandwidth utilization for two systems, %
\emph{IBL} and  \emph{\titleShort{}-ALL}. We observe that \emph{\titleShort{}-ALL} reduces off-chip bandwidth utilization by an average of 17\% compared to \emph{IBL}. This is mainly due to the fact that the larger LLC enabled by \titleShort reduces the number of off-chip main memory requests. To further analyze the reason behind the bandwidth utilization reduction, we measure the LLC MPKI (misses per kilo instructions) for both \emph{IBL} and \emph{\titleShort{}-ALL}. We observe that \emph{\titleShort{}-ALL} reduces the LLC MPKI by 47\% compared to \emph{IBL}.

\subsection{Overhead Analysis}
\label{sec:overhead}
We analyze the storage and power overheads of the additional hardware required by \titleShort{} (i.e., the \titleShort{} controller). \\
\noindent\textbf{Storage cost.} The \titleShort{} controller has two main storage components: the hit/miss prediction unit (\S\ref{sec:misspenalty}) and the extended LLC query unit (\S\ref{sec:warp-status}). The hit/miss prediction unit has 16-KiB of Bloom filter storage per LLC partition. The extended LLC query unit has a 5-KiB on-chip storage per LLC partition to store the request queue, warp status table, and read/write data buffers (\S\ref{sec:warp-status}). Overall, \titleShort{} adds 21 KiB per LLC partition; which is approximately 4\% of the conventional LLC capacity per LLC partition in the NVIDIA RTX 3080 GPU. 

\noindent\textbf{Power Consumption.} We measure the power consumption of the additional hardware required by \titleShort{} using (1)~CACTI~6.5~\cite{muralimanohar2009cacti} for storage units and (2)~the synthesized Verilog HDL models with the NanGate 45nm open cell library~\cite{knudsen2008nangate} for logic units. We observe that the \titleShort{} controller imposes a 0.93\% overhead to the total GPU power consumption. Note that we already take into account \titleShort{}' power consumption overhead for the energy efficiency analysis in \S\ref{sec:energy}.

\section{Related Work}

To our knowledge, this is the first work to propose extending the GPU last-level cache capacity by repurposing the on-chip memory units (i.e., register files, L1 caches, scratchpad memories) of otherwise unused GPU cores. %
In this section, we briefly review  related work in four GPU domains: (1)~\emph{increasing cache capacity}, (2)~\emph{increasing interconnection network performance}, (3)~\emph{controlling cache contention}, and (4)~\emph{helper threads}. 
\\ 
\noindent\textbf{Increasing Cache 
Capacity.} Prior works increase the capacity of the GPU L1 cache~(e.g.,~\cite{jing2016cache,oh2019linebacker,gebhart2012unifying,komuravelli2015stash,darabi2022osm,pekhimenko2012base,sardashti2016yet,sardashti2013decoupled,arelakis2014sc,ghasemazar2020thesaurus,ghasemazar20202dcc,1635958,qureshi2009adaptive,10.1145/3001589,6853233,RNUCA,coop-cache-CPU1,coop-cache-CPU2,7967098,9407080,10.1145/3322127,9138915,7820638}) by %
(1)~utilizing unused registers in the register file~\cite{jing2016cache,oh2019linebacker}, (2)~unifying on-chip memories in the GPU core~\cite{gebhart2012unifying,komuravelli2015stash,darabi2022osm}, (3)~applying cache compression techniques~\cite{pekhimenko2012base,sardashti2016yet,sardashti2013decoupled,arelakis2014sc,ghasemazar2020thesaurus,ghasemazar20202dcc,9138915,8327011}, (4)~caching cooperatively using multiple instances of the L1 cache~\cite{1635958,qureshi2009adaptive,10.1145/3001589,coop-cache-CPU1,coop-cache-CPU2,RNUCA}, and (5) using dense emerging memory technologies, e.g., Domain Wall Memory (DWM)~\cite{6853233}. Since the L1 cache is one of the on-chip memory units that \titleShort repurposes to increase the GPU LLC capacity, we expect these works can further increase the benefits of \titleShort by providing a higher extended LLC capacity. 

Prior works increase the GPU LLC capacity (e.g.,~\cite{samavatian2014efficient,6853233,jing2015energy,samavatian2015architecting,7783731}) by using dense emerging memory technologies, such as STT-MRAM and DWM, to build a conventional LLC with a larger capacity. Compared to these works, \titleShort increases the GPU LLC capacity without (1)~adding any extra on-chip memory, and (2)~depending on emerging memory technologies. %

Some existing GPUs have high LLC capacity. For example, the NVIDIA A100 features a 40-MiB LLC~\cite{choquette2021nvidia}. Although returns diminish, \titleShort can still improve the performance of such a GPU by further increasing the LLC capacity. Since such a large conventional LLC costs significant silicon area (e.g., about 27\% of the total chip area in the A100~\cite{choquette2021nvidia}), \titleShort can help to 1) \emph{reduce} the conventional LLC size in such large-LLC architectures and 2) enable allocating more hardware resources to compute units. 

\noindent\textbf{Improving Interconnection Network Performance.} Prior works (e.g.,~\cite{zhao2016llc,liu2018get,wang2019sharing,mirhosseini2017binochs,mirhosseini2019baran,zhao2016low,zhao2019intra,bakhoda2010throughput,kim2012providing,sadrosadati2017effective,jang2015bandwidth,zhao2020selective,7842945}) increase the GPU LLC throughput by improving interconnection network performance, e.g., by improving network resources~\cite{zhao2016llc,liu2018get,wang2019sharing,mirhosseini2017binochs,mirhosseini2019baran,zhao2016low,zhao2019intra,bakhoda2010throughput,kim2012providing}, better distributing the LLC banks inside the network topologies (e.g., 2D mesh)~\cite{sadrosadati2017effective,jang2015bandwidth}, and duplicating frequently-accessed cache blocks to reduce network contention~\cite{zhao2020selective}. \titleShort relies on the GPU's interconnection network to enable the extended LLC, and therefore, \titleShort' effectiveness can be improved by increasing the performance of the interconnection network. 

\noindent\textbf{Controlling Cache Contention.} Prior works improve the effectiveness of GPU caches by controlling~cache contention (e.g.,~\cite{li2015adaptive,chen2014adaptive,duong2012improving,jia2012characterizing,xie2015coordinated,li2015locality,jia2014mrpb,ausavarungnirun2015exploiting,DBLP:conf/micro/RhuSLE13,wang2018efficient,rogers2012cache,rogers2013divergence,kayiran2013neither,wang2016oaws,li2015priority,mao2015vws,kayiran2014managing,xie2015enabling,10.1145/3037697.3037709}).
These works either (1)~bypass some levels of the cache hierarchy for some memory accesses~\cite{li2015adaptive,chen2014adaptive,duong2012improving,jia2012characterizing,xie2015coordinated,li2015locality,jia2014mrpb,ausavarungnirun2015exploiting,DBLP:conf/micro/RhuSLE13}, or (2)~throttle threads to control cache thrashing~\cite{wang2018efficient,rogers2012cache,rogers2013divergence,kayiran2013neither,wang2016oaws,li2015priority,mao2015vws,kayiran2014managing,xie2015enabling,koo2017access}. \titleShort can be combined with these techniques to enable even higher performance.%

\noindent\textbf{Helper Threads.} Helper threads assist the execution of the main applications threads, by using idle cores or idle cycles for various purposes. Prior works (e.g.,~\cite{darabi2022nura,aamodt2004hardware,brown2002speculative,chappell1999simultaneous,chappell2002microarchitectural,collins2001dynamic,collins2001speculative,dubois2004fighting,dubois1998assisted,ibrahim2003slipstream,kamruzzaman2011inter,kim2002design,lu2005dynamic,luk2001tolerating,mutlu2003runahead,mutlu2005techniques,stephenson2015flexible,zhang2007accelerating,zilles1999use,zilles2001execution,vijaykumar2015case,bauer2011cudadma,DBLP:conf/ics/AamodtC07}) propose helper threads to implement optimization techniques for both CPUs and GPUs, such as data prefetching, pre-computing branch outcomes, managing the caches, or increasing the effective cache bandwidth and capacity by compressing the cache contents. In particular, CABA~\cite{vijaykumar2015case} enables each GPU warp to launch an \emph{assist} warp on the same GPU core to perform data compression, exploiting fine-grained idleness of execution units. \titleShort uses helper threads for a new purpose: extending the capacity of the LLC by exploiting on-chip memories of otherwise unused cores. %

\section{Conclusion}
\label{sec:conclusion}
We introduce \titleShort, the first hardware/software co-designed technique to repurpose otherwise unused GPU cores’ on-chip memories to extend the total GPU last-level cache capacity. \titleShort introduces two execution modes for GPU cores: (1) \computemode, where the core behaves exactly like in a conventional GPU, and (2) \cachemode, where the core lends its on-chip memory space (register file, L1 cache, shared memory) to extend the effective total LLC size. \titleShort{} reuses the on-chip memory resources of a core in \cachemode using a software helper kernel. \titleShort improves the performance and energy efficiency of a baseline NVIDIA RTX~3080 architecture by an average of 39\% and 58\%, respectively, across 14 memory-bound applications. \titleShort performs within 3\% of an ideal baseline with 4$\times$ the LLC, while increasing GPU power consumption by only 0.93\%. We hope that \titleShort{} can help researchers and system designers to rethink how the large on-chip memory resources of GPUs and other accelerators are managed by software and hardware cooperatively. %

\section*{Acknowledgments}
We thank the anonymous reviewers of MICRO 2022 for their encouraging feedback. We thank HPCAN and SAFARI Research Group members for their feedback.  
SAFARI Research Group acknowledges the generous gifts provided by our industrial partners: Google, Huawei, Intel, Microsoft, and VMware. This research was partially supported by the ETH Future Computing Laboratory.

\balance
\bibliographystyle{IEEEtran}
\bibliography{refs}

\end{document}